\documentclass[aps,prd,twocolumn,a4paper,superscriptaddress,nofootinbib]{revtex4-1}
\usepackage{graphicx,caption,subcaption,amsmath,amsfonts,amssymb,multirow,extarrows,bm,acronym,float}
\usepackage[colorlinks,linkcolor=blue,citecolor=blue,urlcolor=blue ]{hyperref}
\usepackage[flushleft]{threeparttable}
\floatstyle{plaintop}
\restylefloat{table}
\usepackage{CJKutf8}
\newcommand{\fs}{\mathcal{F}\text{-statistic}}

\newcommand{\Msun}{\,{\rm M}_\odot}

\newcommand{\DLUT}{School of Physics, Dalian University of Technology, Dalian 116024, People's Republic of China}

\begin{document}

\title{A Robust and Efficient $\mathcal{F}$-statistic-based Framework for Consistent Bayesian Inference of Compact Binary Coalescences}

\author{Hai-Tian Wang (\begin{CJK}{UTF8}{gbsn}王海天\end{CJK})}
\email{wanght9@dlut.edu.cn}
\affiliation{\DLUT}

\date{\today}

\begin{abstract}
We present a comprehensive investigation of the $\fs$ method for parameter estimation of gravitational wave (GW) signals from compact binary coalescences. By analytically maximizing the likelihood over the luminosity distance and polarization angle, this approach reduces the dimensionality of the parameter space to enhance computational efficiency. We also introduce a novel formulation for calculating the Bayesian evidence for the $\fs$, enabling a quantitative assessment of its performance against standard full frequency-domain (FFD) Bayesian inference. Applying these two methods to analyze several representative GW events (GW190412, GW190814, and GW170817), we find that the $\fs$ consistently yields results in good agreement with the FFD approach, while offering a significant reduction in computational cost. We demonstrate that including calibration uncertainty generally improves the agreement between the two methods. Furthermore, under the assumption of physical priors, the $\fs$-based analyses consistently yield higher Bayesian evidence than the corresponding FFD analyses. While the $\fs$ produces slightly broader constraints on some parameters, we argue this represents a more honest uncertainty quantification, particularly in high-dimensional parameter spaces with complex posterior structures. These results highlight the significant advantages of the $\fs$ method for GW data analysis, positioning it as a powerful tool for the era of high-rate detections with future observatories.
\end{abstract}

\maketitle

\acrodef{GW}{gravitational wave}
\acrodef{LIGO}{Laser Interferometer Gravitational-Wave Observatory}
\acrodef{LVK}{LIGO-Virgo-KAGRA}
\acrodef{LVKC}{LIGO-Virgo-KAGRA Collaboration}
\acrodef{FFD}{full frequency-domain}
\acrodef{FIM}{Fisher Information Matrix}
\acrodef{NR}{numerical relativity}
\acrodef{TD}{time-domain}
\acrodef{BH}{black hole}
\acrodef{BBH}{binary black hole}
\acrodef{NSBH}{neutron star-black hole}
\acrodef{BNS}{binary neutron star}
\acrodef{GR}{general relativity}
\acrodef{PN}{post-Newtonian}
\acrodef{SNR}{signal-to-noise ratio}
\acrodef{PSD}{power spectral density}
\acrodef{PDF}{probability density function}
\acrodef{ACF}{auto-covariance function}
\acrodef{JSD}{Jensen-Shannon divergence}
\acrodef{IMR}{inspiral-merger-ringdown}

\section{Introduction}\label{sec:intro}

During the initial three observing runs conducted by the \ac{LVKC}, over $90$ gravitational wave events were identified \citep{LIGO_PRX2019, LIGO_O3a_PRX2020, KAGRA:2021vkt}. 
Notably, more than $90\%$ of these detections are \ac{BBH} mergers, making them the dominant source type in the \ac{GW} catalog. 
These \ac{BBH} systems are invaluable for testing \ac{GR}, as their strong-field dynamics offer stringent constraints on deviations from Einstein's predictions \citep{LIGOScientific:2016lio,LIGOScientific:2020tif,LIGOScientific:2021sio}.

\ac{GW} signals are buried deep within detector noise, requiring sophisticated methods to extract their underlying physical parameters. 
Researchers typically employ Bayesian inference for this purpose. 
Although conventional Bayesian methods have been successfully applied in both real data analysis and simulated signal injection tests \citep{Veitch:2014wba,Romero-Shaw:2020owr,Ashton:2018jfp,Biwer:2018osg}, they face significant challenges when analyzing the majority of detected \ac{GW} events, which often have low \ac{SNR}. 
Furthermore, systematic uncertainties arising from detector calibration can bias the recovery of physical parameters if not properly accounted for \citep{LIGO-T1400682,Vitale:2011wu,Payne:2020myg,Sun:2020wke}.
For such events, the high-dimensional parameter space (up to $15$ dimensions) leads to prolonged computational times for parameter estimation. 
Moreover, when the posterior distribution exhibits a multimodal structure, the accuracy of parameter estimation can diminish. 
These limitations highlight the need for more efficient and robust approaches in \ac{GW} data analysis.

To address these challenges, researchers have developed various techniques to enhance the standard Bayesian framework, such as marginalization \citep{Thrane:2018qnx,Hoy:2022tst,Roulet:2024hwz}, reduced order quadrature algorithms \citep{Antil:2012wf,Smith:2016qas,Smith:2021bqc}, machine learning \citep{Cranmer:2019eaq,Dax:2022pxd,Wong:2023lgb,Marx:2024wjt}, and the $\fs$ \citep{Kokkotas:1994ef,Jaranowski:1998qm,Wang:2024jlz}. 
The core innovation of the $\fs$ is its analytical maximization over certain time- and frequency-independent parameters, which reduces the dimensionality of the parameter space and enhances both the efficiency and accuracy of parameter estimation. 
The $\fs$ method has been widely applied in \ac{GW} research, including searches for continuous waves from isolated neutron stars \cite{KAGRA:2022dwb}, binary white dwarf systems \cite{Prix:2007zh}, and \acp{BBH} \cite{Wang:2014ava}; investigations of extreme mass-ratio inspirals \cite{Wang:2012xh}; and studies of ringdown signals \citep{Wang:2024jlz,Wang:2024yhb,Tao:2024oyx} or the full \ac{IMR} signal \citep{Harry:2010fr,Keppel:2012ye,Harry:2016ijz}.

Previous applications of the $\fs$ in the context of \ac{IMR} \ac{GW} analyses have largely been developed with a focus on signal searches and detection studies. These early investigations typically employed waveform models appropriate to the computational and modeling capabilities available at the time, and were instrumental in establishing the feasibility and utility of the $\fs$ framework for \ac{GW} data analysis \cite{Harry:2010fr}.
Since then, \ac{GW} modeling and inference techniques have undergone substantial advances. Contemporary \ac{IMR} analyses routinely employ waveform families that incorporate higher-order modes and orbital precession, such as \texttt{IMRPhenomXPHM} \cite{Pratten:2020ceb} and \texttt{IMRPhenomXPHM-SpinTaylor} \cite{Colleoni:2024knd}, which are now known to play an important role in accurately describing events like GW190412 \cite{LIGOScientific:2020stg} and GW190814 \cite{LIGOScientific:2020zkf}. In parallel, Bayesian parameter estimation methods have matured significantly, supported by increasingly sophisticated sampling algorithms.
Motivated by these developments, this work revisits the $\fs$ method in the context of modern, physically complete waveform models and focuses on its application to parameter estimation for individual \ac{IMR} events. We further introduce a consistent formulation for computing the Bayesian evidence associated with the $\fs$, which enables direct and quantitative comparisons with standard Bayesian inference approaches. Our results demonstrate that, in this setting, the $\fs$ method provides a computationally efficient and robust alternative for \ac{GW} parameter estimation.

This paper is organized as follows. In Sec.~\ref{sec:methods}, we detail the framework for \ac{FFD} Bayesian inference and the derivation of the $\fs$ method. We then present the results of our comparative analysis using data from different types of events in Sec.~\ref{sec:bayes}. Finally, we provide a summary of our findings and discuss their implications in Sec.~\ref{sec:con}. Unless stated otherwise, we use geometric units where $G=c=1$ throughout this paper. 

\section{Methodology}\label{sec:methods}
In this section, the statistical methods employed for parameter estimation are presented. We begin by outlining the conventional \ac{FFD} framework, which serves as the benchmark for our analysis. Subsequently, we detail the derivation of the $\fs$ for a complete \ac{IMR} signal. This derivation proceeds in two stages: first, the waveform model is reformulated to isolate parameters that appear linearly in the signal model; second, the profile likelihood is constructed by analytically maximizing the likelihood function with respect to these parameters. Following the derivation, we describe the procedure for reconstructing the posterior distributions of the analytically maximized parameters. A novel formulation for calculating the Bayesian evidence, and consequently the Bayes factor, for the $\fs$ method is also introduced. The section concludes with an outline of the prior distributions and the specific sampler configurations adopted for this work.

\subsection{Full Bayesian Inference Framework}\label{ssec:fbi}
In Bayesian inference, the posterior \ac{PDF} of a set of parameters $\Theta$, given the data $\textbf{d}$, is determined by Bayes' theorem:
\begin{equation}
P(\Theta|\textbf{d}) = \frac{\mathcal{L}(\textbf{d}|\Theta) P(\Theta)}{P(\textbf{d})},
\end{equation}
where $\mathcal{L}(\textbf{d}|\Theta)$ is the likelihood function, $P(\Theta)$ is the prior \ac{PDF} representing our knowledge before analyzing the data, and $P(\textbf{d})$ is the evidence. The evidence is a normalization constant for parameter estimation but is crucial for model comparison.

For a \ac{GW} signal $\textbf{h}(\Theta)$ embedded in stationary and Gaussian noise, the likelihood is constructed using the noise-weighted inner product, defined as:
\[
\langle \textbf{h}_{1}(f)|\textbf{h}_{2}(f)\rangle = 4\int_{f_{\text{low}}}^{f_{\text{high}}}\frac{\textbf{h}_{1}^{*}(f)\textbf{h}_{2}(f)}{S_{n}(f)}\mathrm{\textbf{d}}f,
\]
where $\textbf{h}_{1,2}(f)$ are frequency-domain waveforms or data, $\textbf{h}^*$ denotes complex conjugation, and $S_{n}(f)$ is the one-sided \ac{PSD} of the detector noise. The log-likelihood function, $\ln\mathcal{L}$, is then given by
\begin{equation}
\begin{aligned}\label{eq:ll0}
\ln\mathcal{L}(\Theta)&=-\frac{1}{2}\langle \textbf{d}-\textbf{h}(\Theta)|\textbf{d}-\textbf{h}(\Theta)\rangle\\
&=\ln\Lambda(\Theta)-\frac{1}{2}\langle \textbf{d}|\textbf{d}\rangle,
\end{aligned}
\end{equation}
where the log-likelihood ratio, $\ln\Lambda(\Theta)$, is
\begin{equation}\label{eq:ll1}
\ln\Lambda(\Theta)=\langle \textbf{d}|\textbf{h}(\Theta)\rangle-\frac{1}{2}\langle \textbf{h}(\Theta)|\textbf{h}(\Theta)\rangle.
\end{equation}
This approach requires numerical sampling over the \textit{full} set of model parameters, $\Theta$. Throughout this paper, we refer to this method as \ac{FFD}.

\subsection{The $\mathcal{F}$-statistic Method}
The $\fs$ method offers an alternative to the \ac{FFD} approach by reducing the dimensionality of the parameter space, thereby increasing computational efficiency. This is achieved by analytically maximizing the likelihood function over a subset of parameters that appear linearly in the waveform model. The derivation consists of two primary stages, which are detailed in the following subsections. First, the waveform template is reformulated to isolate these ``linear" parameters from the remaining ``non-linear" parameters, $\bm\theta$. Second, the profile likelihood, known as the $\fs$, is constructed by analytically solving for the values of the linear parameters that maximize the likelihood for any given set of the non-linear parameters.

\subsubsection{Waveform Reformulation}
A frequency-domain \ac{IMR} signal from a \ac{BBH} merger is composed of two polarization modes, $\textbf{h}_{+}(f)$ and $\textbf{h}_{\times}(f)$. The signal observed in a detector, $\textbf{h}(f)$, is a linear combination of these modes, weighted by the antenna pattern functions, $F_+$ and $F_\times$:
\begin{equation}
\textbf{h}(f)=F_{+}\textbf{h}_{+}(f)+F_{\times}\textbf{h}_{\times}(f).
\end{equation}
The antenna patterns are functions of the source's sky location (right ascension $\alpha$, declination $\delta$) and the polarization angle $\psi$. Specifically, they take the form:
\begin{equation}
\begin{aligned}
F_{+}&=F_{1}\cos2\psi - F_{2}\sin2\psi,\\
F_{\times}&=F_{1}\sin2\psi + F_{2}\cos2\psi,
\end{aligned}
\end{equation}
where $F_{1,2}$ are functions of $\alpha$ and $\delta$. For the short-duration signals typical of \ac{LVKC} detections, the time-variation of the antenna patterns is negligible, and they can be treated as independent of frequency.

The full waveform model depends on a total of $15$ intrinsic and extrinsic parameters. The central idea of the $\fs$ is to reformulate the waveform to separate a subset of these parameters—the ``linear" parameters—from the remaining ``non-linear" parameters, denoted by the vector $\bm\theta$. The waveform is thus written as a linear combination of basis waveforms $\textbf{h}_\mu$:
\begin{equation}
\textbf{h}(f) = B^{\mu}\textbf{h}_{\mu}(f),
\end{equation}
where the Einstein summation convention is implied\footnote{In this work, we adopt the Einstein summation convention.}. The coefficients $B^{\mu}$ are functions of the linear parameters, while the basis waveforms $\textbf{h}_\mu(f)$ depend on $\bm\theta$.

Ideally, four parameters are candidates for this linear separation: luminosity distance ($d_L$), polarization angle ($\psi$), orbital inclination ($\iota$), and a reference phase ($\phi_c$). However, for realistic waveform models that include higher-order multipoles and orbital precession, the dependencies on $\iota$ and $\phi_c$ become highly complex and are entangled with the non-linear parameters. Therefore, in practice, only $d_L$ and $\psi$ can be analytically isolated.

By substituting the expressions for the antenna patterns into the detector response and separating the terms containing $d_L$ and $\psi$, the waveform can be expressed as $h(f)=B^{1}\textbf{h}_{1}(f)+B^{2}\textbf{h}_{2}(f)$, where:
\begin{equation}
\begin{aligned}
B^{1}&=\frac{\cos2\psi}{d_{L}}, \\
B^{2}&=\frac{\sin2\psi}{d_{L}},\\
\textbf{h}_{1}(f)&=F_{1}\textbf{h}_{+}(f)+F_{2}\textbf{h}_{\times}(f),\\
\textbf{h}_{2}(f)&=F_{1}\textbf{h}_{\times}(f)-F_{2}\textbf{h}_{+}(f).
\end{aligned}\label{eq:b12}
\end{equation}
This reformulation is the first essential step toward constructing the $\fs$.

\subsubsection{Deriving the Profile Likelihood}
By substituting the reformulated waveform from Eq.~\eqref{eq:b12} into the log-likelihood ratio of Eq.~\eqref{eq:ll1}, the likelihood can be expressed in terms of the linear parameters $B^{\mu}$ and non-linear parameters $\bm\theta$:
\begin{equation}\label{eq:ll2}
\ln\Lambda(\Theta)=B^{\mu}\textbf{s}_{\mu}(\bm\theta)-\frac{1}{2}B^{\mu}\textbf{M}_{\mu\nu}(\bm\theta)B^{\nu},
\end{equation}
where $\textbf{s}_{\mu} \equiv \langle \textbf{d}|\textbf{h}_{\mu}\rangle$ is the projection of the data onto the basis waveforms, and $\textbf{M}_{\mu\nu} \equiv \langle \textbf{h}_{\mu}|\textbf{h}_{\nu}\rangle$ is the covariance matrix of these basis waveforms.

To derive the profile likelihood, we analytically maximize this expression with respect to $B^{\mu}$. This is achieved by setting the partial derivative of $\ln\Lambda(\Theta)$ with respect to $B^\nu$ to zero:
\begin{equation}\label{eq:par2}
\frac{\partial\ln\Lambda(\Theta)}{\partial B^{\nu}}=\textbf{s}_{\nu}-B^{\mu}\textbf{M}_{\mu\nu}=0.
\end{equation}
Solving for $B^{\mu}$ yields the maximum likelihood estimator, $\hat{B}^{\mu}$:
\begin{equation}\label{eq:B}
\hat{B}^{\mu}=\textbf{M}^{\mu\nu}\textbf{s}_{\nu},
\end{equation}
where $\textbf{M}^{\mu\nu}$ is the inverse of the matrix $\textbf{M}_{\mu\nu}$. The matrix of second partial derivatives is negative-definite, ensuring that $\hat{B}^{\mu}$ corresponds to a unique maximum for any given $\bm\theta$. Note that the values of $\hat{B}^{\mu}$ do not depend on the priors of $B^{\mu}$.

Following \citet{Wang:2024yhb}, substituting this result back into Eq.~\eqref{eq:ll2} allows the log-likelihood ratio to be rewritten as:
\begin{equation}\label{eq:fs0}
\ln\Lambda(\Theta)=\frac{1}{2}\textbf{s}_{\mu}\textbf{M}^{\mu\nu}\textbf{s}_{\nu}-\frac{1}{2}(B^{\mu}-\hat{B}^{\mu})\textbf{M}_{\mu\nu}(B^{\nu}-\hat{B}^{\nu}).
\end{equation}
This rearranged form is powerful because, according to the conjunction rule of probability, $P(\bm\theta,B^{\mu}|\textbf{d})=P(\bm\theta|\textbf{d})P(B^{\mu}|\bm\theta,\textbf{d})$, it separates the posterior into two distinct components:
\begin{align}
P(\bm\theta|\textbf{d})&\propto e^{\frac{1}{2}\textbf{s}_{\mu}\textbf{M}^{\mu\nu}\textbf{s}_{\nu}}p(\bm\theta),\label{eq:pp1}\\
P(B^{\mu}|\bm\theta,\textbf{d})&\propto e^{-\frac{1}{2}(B^{\mu}-\hat{B}^{\mu})\textbf{M}_{\mu\nu}(B^{\nu}-\hat{B}^{\nu})}p(B^{\mu}|\bm\theta),\label{eq:pp2}
\end{align}
where $p(\bm\theta)$ and $p(B^{\mu}|\bm\theta)$ are the prior distributions.

From this separation, we can define the log-likelihood for the non-linear parameters $\bm\theta$ as:
\begin{align}
\ln\Lambda(\bm\theta)&= \frac{1}{2}\textbf{s}_{\mu}\textbf{M}^{\mu\nu}\textbf{s}_{\nu},\label{eq:fs1}
\end{align}
which is defined as the $\fs$ and is the log-profile likelihood used in this study for sampling $\bm\theta$. The log-likelihood for the conditional posterior of $B^{\mu}$ is:
\begin{align}
\ln\Lambda(B^{\mu}|\bm\theta)\footnotemark[2] &= -\frac{1}{2}(B^{\mu}-\hat{B}^{\mu})\textbf{M}_{\mu\nu}(B^{\nu}-\hat{B}^{\nu}).\label{eq:ll_b}
\end{align}
For an analysis involving multiple detectors, the network $\fs$ is obtained by summing the $\textbf{s}_\mu$ and $\textbf{M}_{\mu\nu}$ terms from each detector.
\footnotetext[2]{It is written in this form for simplicity but is not rigorously formal; the complete expression should be written as $P(B^{\mu}|\bm\theta,\textbf{d})/p(B^{\mu}|\bm\theta)$.}
\setcounter{footnote}{2}

\subsection{Posterior Reconstruction for Maximized Parameters}\label{ssec:plp}
Once the likelihood is expressed in the separated form of Eq.~\eqref{eq:fs0}, there are two distinct statistical approaches for parameter estimation. The first approach is to use Eq.~\eqref{eq:fs0} as the likelihood and marginalize it over $B^{\mu}$ for sampling. This method has been adopted in several studies~\citep{Whelan:2013xka,Dong:2025igh} and, importantly, has been shown to produce posterior distributions consistent with the standard \ac{FFD} approach for high \ac{SNR} cases \citep{Dong:2025igh}.
The second approach, which we adopt in this work, leverages the probabilistic decomposition shown in Eqs.~\eqref{eq:pp1} and \eqref{eq:pp2}. By using only the profile likelihood (the $\fs$ from Eq.~\eqref{eq:fs1}) for the primary analysis, we sample only the space of the non-linear parameters $\bm\theta$. This reduction in the dimensionality of the sampled parameter space is the key to the enhanced computational efficiency of the method.

A consequence of this approach is that the posterior distributions for the linear parameters $B^{\mu}$ (and thus for luminosity distance $d_L$ and polarization angle $\psi$) are not obtained directly from the sampler. Instead, they must be constructed in a post-processing step. The $\fs$ analysis yields $N_s$ posterior samples, $\{\boldsymbol{\theta}_i\}$, drawn from $P(\boldsymbol{\theta} | \textbf{d})$. For each sample $\boldsymbol{\theta}_i$, the conditional posterior for $B^{\mu}$, given by Eq.~\eqref{eq:pp2} and assuming a uniform prior $p(B^{\mu} | \boldsymbol{\theta})$, is a multivariate Gaussian distribution:
\begin{equation}
P(B^{\mu} | \boldsymbol{\theta}_i, \textbf{d}) \sim \mathcal{N}\left(\hat{B}^\mu(\boldsymbol{\theta}_i), \textbf{M}^{\mu\nu}(\boldsymbol{\theta}_i)\right),
\end{equation}
where the mean vector $\hat{B}^\mu(\boldsymbol{\theta}_i)$ and covariance matrix $\textbf{M}^{\mu\nu}(\boldsymbol{\theta}_i)$ are calculated using the specific values from the sample $\boldsymbol{\theta}_i$. 
To obtain the full marginal posterior $P(B^{\mu} | \textbf{d})$, we marginalize over the uncertainty in $\bm\theta$ via composition: for each posterior sample $\boldsymbol{\theta}_i$, we draw $500$ corresponding samples of $B^{\mu}$ from its conditional Gaussian distribution.

This procedure initially yields posteriors corresponding to a uniform prior on $B^{\mu}$. To apply a different, physically motivated target prior, we use importance resampling. The weight for each sample is the ratio of the target prior to the original sampling prior:
\begin{equation}\label{eq:iw1}
w_i = \frac{p(B^{\mu}_i | \boldsymbol{\theta}_i, \otimes)}{p(B^{\mu}_i | \boldsymbol{\theta}_i,\ominus)},
\end{equation}
where $\ominus$ and $\otimes$ denote the uniform and target priors on $B^{\mu}$, respectively.

\subsection{Priors}
In our framework, the non-linear parameters $\bm\theta$ and the linear parameters $B^{\mu}$ are assumed to be independent in the prior distribution. The joint prior therefore factorizes as $P(\Theta) = p(\bm\theta) p(B^{\mu})$, and the conditional prior $p(B^{\mu}|\bm\theta)$ simplifies to $p(B^{\mu})$.

The reconstruction method described in Sec.~\ref{ssec:plp} most naturally operates with a uniform prior on the linear parameters $B^{\mu}$. It is crucial, however, to understand the prior this implies for the physical parameters: the luminosity distance $d_L$ and the polarization angle $\psi$. The relationship is described by the transformation:
\begin{equation}\label{eq:pb}
p(B^{\mu})=d_L^3p(d_L,\psi).
\end{equation}
A uniform prior on $B^{\mu}$ indicates a uniform prior on $\psi \in [0, \pi)$ and an unphysical prior on the luminosity distance, where $p(d_L) \propto 1/d_L^3$. The physically motivated prior, which we use as our target prior for reporting final results, assumes sources are distributed uniformly in comoving volume. 
For the non-linear parameters $\bm\theta$, we adopt the same priors as those used by the \ac{LVKC} in their catalog analyses~\citep{LIGO_PRX2019,LIGO_O3a_PRX2020,KAGRA:2021vkt}.

\subsection{Evidence and Bayes Factor Calculation}
A key metric for model comparison is the Bayesian evidence, $\mathcal{Z}$. For the \ac{FFD} method, the evidence for the full parameter set $\Theta$ is a direct output of nested sampling algorithms:
\begin{equation}\label{eq:z0}
\mathcal{Z}(\Theta) = \int p(\bm\theta, B^{\mu}) \mathcal{L}(\bm\theta, B^{\mu}) \textbf{d}\bm\theta dB^{\mu}.
\end{equation}
In contrast, the $\fs$ method, by sampling only the non-linear parameters $\bm\theta$, directly yields only the evidence for that reduced parameter space, $\mathcal{Z}(\bm\theta)=\int p(\bm\theta) \mathcal{L}(\bm\theta) \textbf{d}\bm\theta$. 
Similar to \citet{Wang:2025rvn}, we propose a rigorous method for incorporating the prior of the maximized parameters, $B^{\mu}$, into the evidence calculation for the $\fs$.

Specifically, we introduce a novel formulation to compute the full evidence for the $\fs$ framework, thereby enabling a robust Bayes factor comparison. The full evidence under a target prior (denoted by $\otimes$) is
\begin{align}
\mathcal{Z}(\Theta|\otimes) &= \int \mathcal{L}(\bm\theta, B^{\mu}) p(\bm\theta, B^{\mu} | \otimes) \textbf{d}\bm\theta dB^{\mu} \notag\\
&= \int \mathcal{L}(\bm\theta) \left( \int \Lambda(B^{\mu} | \bm\theta) p(B^{\mu} | \bm\theta, \otimes) dB^{\mu} \right) p(\bm\theta) \textbf{d}\bm\theta \notag\\
&= \int f(\bm\theta) \mathcal{L}(\bm\theta) p(\bm\theta) \textbf{d}\bm\theta.
\end{align}
This integral can be approximated as an expectation value over the posterior samples of $\bm\theta$:
\begin{equation}
\mathcal{Z}(\Theta|\otimes) \approx \frac{\mathcal{Z}(\bm\theta)}{N(\bm\theta)} \sum_{i=1}^{N(\bm\theta)} f(\bm\theta_i),\label{eq:z1}
\end{equation}
where the sum is over the $N(\bm\theta)$ posterior samples of $\bm\theta$. The function $f(\bm\theta)$ is calculated for each sample $\bm\theta_i$ as follows:
\begin{align}
f(\bm\theta) &= \int \Lambda(B^{\mu} | {\bm\theta}) p(B^{\mu} | {\bm\theta}, \otimes) dB^{\mu} \notag\\
&\approx \frac{\mathcal{Z}(B^{\mu}|{\bm\theta}, \ominus)}{N(B^{\mu}|{\bm\theta}, \ominus)} \sum_{j=1}^{N(B^{\mu}|{\bm\theta}, \ominus)} \frac{p(B^{\mu}_j | {\bm\theta}, \otimes)}{p(B^{\mu}_j | {\bm\theta}, \ominus)}.\label{eq:f1}
\end{align}
In this expression, the sum is over $N(B^{\mu}|{\bm\theta}, \ominus)$ samples (we use $500$ in this study) drawn from the conditional posterior $P(B^{\mu} | {\bm\theta}, \textbf{d}, \ominus)$, which corresponds to a simple uniform prior (denoted by $\ominus$). $\mathcal{Z}(B^{\mu}|{\bm\theta}, \ominus)$ is the evidence for the linear parameters under this uniform prior. With the full evidence $\mathcal{Z}_{\mathrm Fs}$ computed, the Bayes factor, $\mathcal{B}^{\rm s1}_{\rm s2} = \mathcal{Z}_{\rm s1} / \mathcal{Z}_{\rm s2}$, can be used to quantitatively assess the stability under different sampler configurations.

It is important to emphasize that this evidence comparison is strictly conditional on the assumption that the joint prior factorizes as $P(\Theta) = p(\bm{\theta}) p(B^{\mu})$, which implies $p(B^{\mu} \mid \bm{\theta}) = p(B^{\mu})$. This assumption is satisfied in essentially all current LVK IMR analyses relevant to this work. Under this condition, the $\fs$ evidence computed via importance sampling with physically motivated priors can in principle be compared directly with the FFD evidence. However, if one adopts a prior where the luminosity distance is not independent of other parameters (for example, a prior jointly coupled to the inclination angle), the current $\fs$ evidence expression would require modification, and a direct comparison would no longer be automatic. Furthermore, while a systematically higher $\fs$ evidence is qualitatively consistent with higher likelihood values, we caution against overstating this result; Bayesian evidence is most safely used for Bayes-factor-based model comparison within the same inference framework. We note that consistency between Bayes factors computed in the $\fs$ framework and in full Bayesian inference has already been examined in previous ringdown studies \citep{Wang:2024yhb,Wang:2025rvn,Wang:2025baj}.

\subsection{Sampler Configuration}
For the main analyses, Bayesian inference was performed using the {\sc Bilby} package~\citep[v2.7.0;][]{Ashton:2018jfp}, with the nested sampling algorithm implemented in {\sc Dynesty}~\citep[v3.0.0;][]{Romero-Shaw:2020owr}.
We adopted a robust sampler configuration: $1000$ live points were used, the primary sampling method was set to ``rwalk'', the number of autocorrelation lengths was fixed to $50$, and new points were selected using the ``live'' method—settings consistent with those employed in the GWTC-2.1 \citep{LIGO_O3a_PRX2020} and GWTC-3 \citep{KAGRA:2021vkt} catalog analyses. For completeness, results obtained with alternative configurations—including different versions of {\sc Bilby} and {\sc Dynesty}, as well as varied sampler settings—are presented in the Appendix; all yield consistent conclusions.
All analyses were parallelized using $128$ threads to reduce computational time.

\section{Analysis and Results}\label{sec:bayes}
In this work, we present a detailed comparison of parameter estimation results using the $\fs$ and FFD methods across three selected \ac{GW} events: GW170817 \citep{gw170817_PRL2017}, GW190412 \citep{LIGOScientific:2020stg}, and GW190814 \citep{LIGOScientific:2020zkf}. 
Notably, GW190814 has a mass ratio close to $10$, with the secondary compact object's mass around $2.6\,M_\odot$, leaving open the possibility that it is a \ac{NSBH} system. 
Therefore, for GW190814 we perform independent analyses using two different waveform models: \texttt{IMRPhenomXPHM-SpinTaylor} (a general-purpose model for black-hole binary coalescences) and \texttt{IMRPhenomNSBH} (specifically designed for \ac{NSBH} systems).
For completeness and as an additional benchmark, the analysis of the classic \ac{BBH} event GW150914 is included in the Appendix.~\ref{apsec:gw150914}.

Taken together, the test set covers several representative classes of \ac{GW} sources currently reported by \ac{LVKC}:
\begin{itemize}
	\item \Ac{BBH} systems: GW190412 \& GW190814 (where both spin precession and higher harmonics contribute significantly), both analyzed with \texttt{IMRPhenomXPHM-SpinTaylor};
	\item \Ac{NSBH} system: GW190814, analyzed with \texttt{IMRPhenomNSBH};
	\item \Ac{BNS} system: GW170817, analyzed with \texttt{IMRPhenomXP-NRTidalv3} (including tidal deformability effects).
\end{itemize}
Note that our $\fs$ workflow is fully compatible with calibration error: the additional parameters induced by calibration uncertainty follow the standard spline model used in LVKC analyses \citep{LIGO-T1400682,Vitale:2011wu,Payne:2020myg,Sun:2020wke}. These parameters enter the waveform model in a non-linear manner and are therefore treated in the same way as other non-linear parameters. 
Therefore, we present the analysis results both with and without taking into account the effect of calibration uncertainties. 

For each source type, analysis settings---including parameter priors, sampling rate, data segment length, and other configuration choices---have been chosen to closely match the original \ac{LVKC} published analyses, ensuring fair and meaningful comparison. 
Detailed configuration parameters for each event are provided together with the posterior samples accompanying this work, allowing full reproducibility of the presented results \cite{DataFsIMR150914}.

\begin{figure*}
\centering
\begin{subfigure}[b]{0.48\linewidth}
\centering
\includegraphics[width=\textwidth,height=6.6cm]{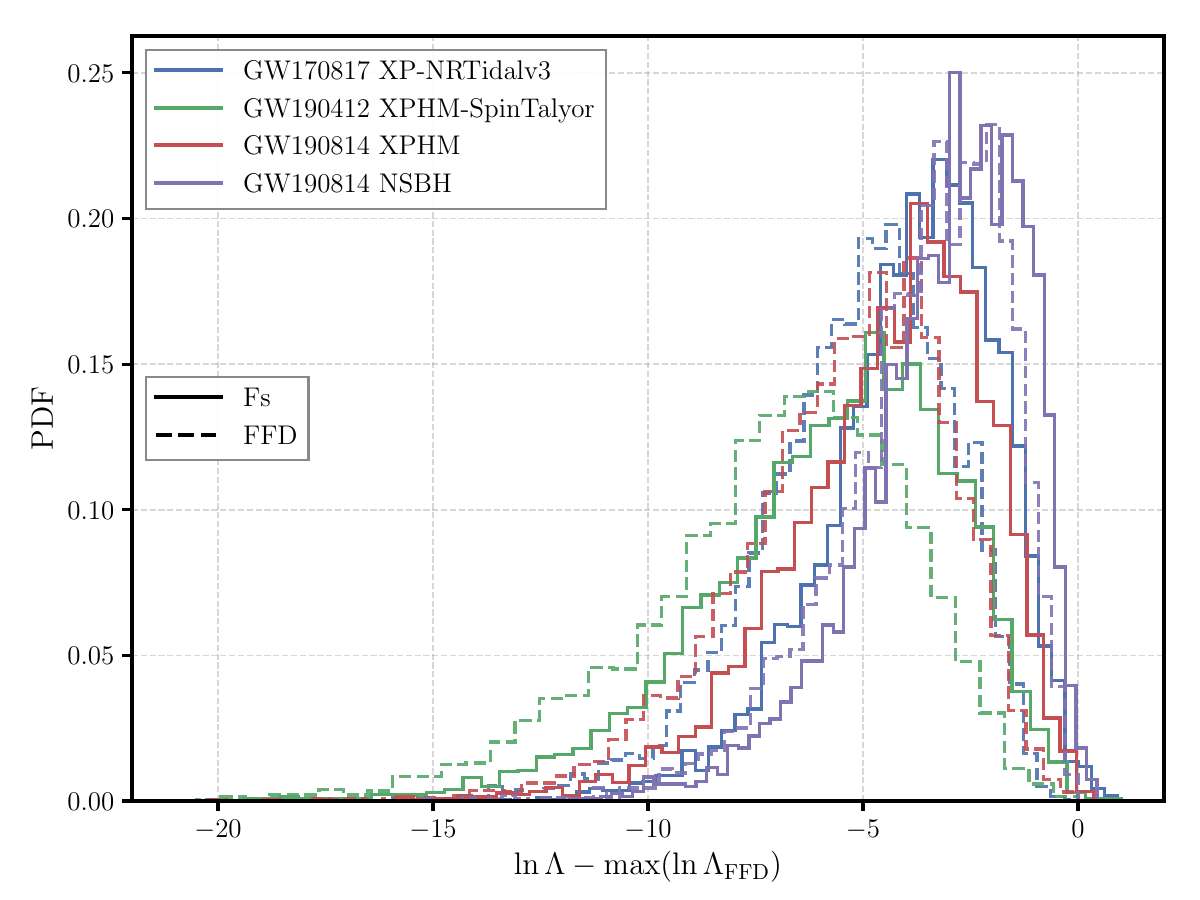}
\end{subfigure}%
\begin{subfigure}[b]{0.48\linewidth}
\centering
\includegraphics[width=\textwidth,height=6.6cm]{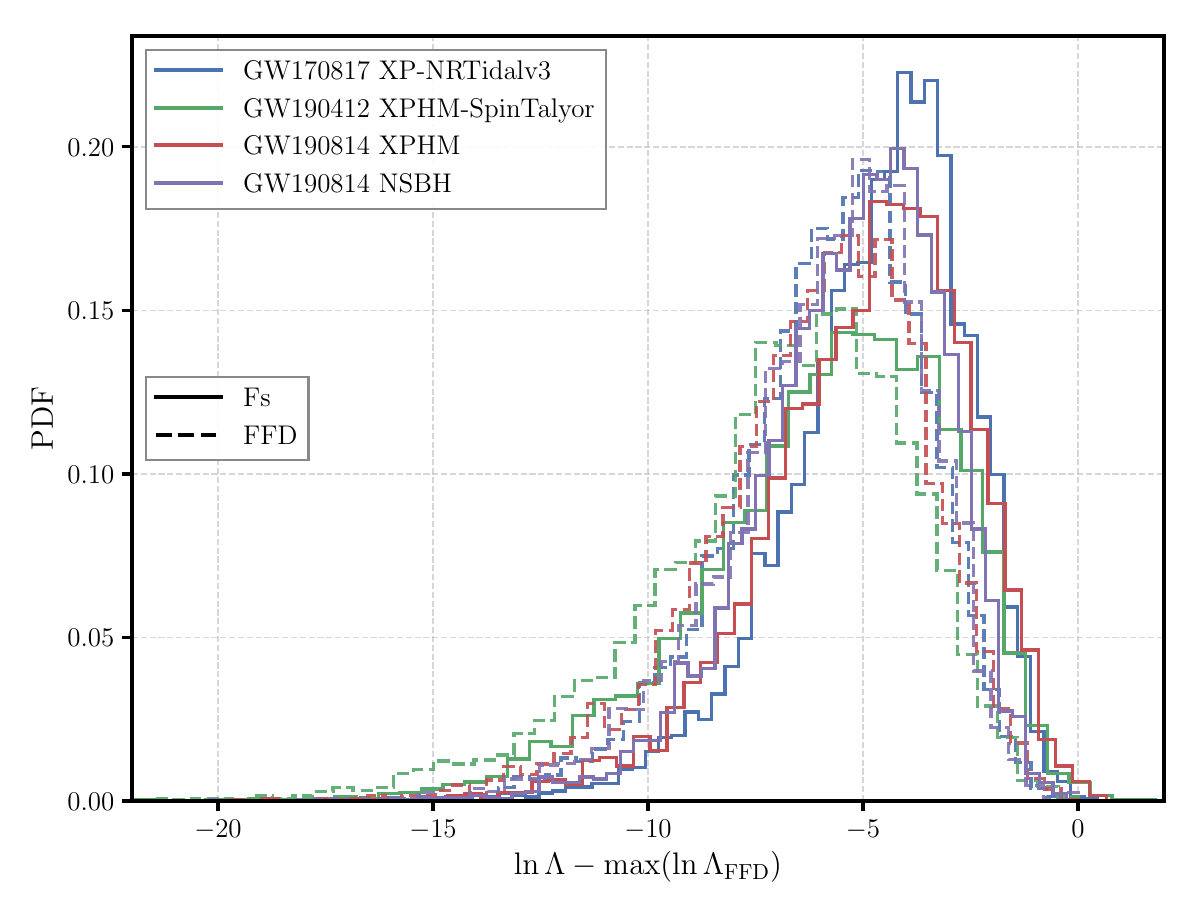}
\end{subfigure}%
\caption{
Posterior distributions of the normalized log-likelihood ratio, $\ln \Lambda - \max(\ln \Lambda_{\mathrm{FFD}})$, for a set of representative gravitational-wave events, where the normalization is performed using the maximum log-likelihood obtained in the corresponding \ac{FFD} analysis for each event.
The dashed histograms show results from the \ac{FFD} method, while the solid histograms correspond to the $\fs$ method (labeled as ``Fs'').
Different colors denote different events, as indicated in the legend.
The left panel presents results obtained without accounting for calibration uncertainties, whereas the right panel shows results obtained when calibration uncertainty parameters are included and propagated through the inference via nested sampling.
Interestingly, incorporating calibration uncertainties has a relatively minor impact on the differences between the two methods for a given event, while it significantly reduces the variation of the log-likelihood distributions across different events, leading to a greater degree of consistency among event-specific results.
}
\label{fig:llall}
\end{figure*}

Our analysis is structured as follows. We begin by evaluating the overall performance and efficiency of the two methods by comparing the distributions of the log-likelihood ratio, $\ln\Lambda$, shown in Fig.~\ref{fig:llall}. 
In each comparison between the \ac{FFD} method and the $\fs$ method across different cases, the distributions associated with the $\fs$ method are consistently shifted toward higher values. This indicates a better fit to the data, a result attributable to the optimization of two linear parameters.
For a more definitive and quantitative comparison, we compute the evidence, $\mathcal{Z}$, using the procedure outlined in Sec.~\ref{sec:methods}.

Next, we present results of each set of analyses one by one.
For each analysis, we examine the posterior distributions of the luminosity distance ($d_L$) and polarization angle ($\psi$), which are analytically maximized in the $\fs$ framework and subsequently reconstructed. Furthermore, we provide a comprehensive comparison of all remaining source parameters and assess the stability of each method across different sampler configurations using the \ac{JSD} (see Ref.~\cite{Wang:2024liy} for details).

\subsection{Analysis of GW190412: An Asymmetric-Mass, Precessing Binary}
We next apply our comparative framework to the \ac{BBH} event GW190412 \citep{LIGOScientific:2020stg}. As a representative example of a complex merger alongside GW190814, this event features a highly asymmetric mass ratio, spin-induced orbital precession, and significant contributions from higher-order radiation multipoles. These physical properties make GW190412 an excellent test case for inference methods. In particular, the presence of higher-order modes helps break degeneracies involving extrinsic parameters, improving constraints on the luminosity distance and polarization angle. Because these are precisely the parameters that the $\fs$ analytically maximizes over, this event provides a stringent test of the method's ability to accurately reconstruct their posterior distributions while accelerating the analysis.

For GW190412, we conduct four parallel analyses, applying both the \ac{FFD} and $\fs$ methods with and without marginalization over calibration uncertainties. All other settings are held identical to ensure a controlled comparison. This subsection first focuses on the inferred posterior distributions for the analytically maximized parameters—luminosity distance and polarization angle—before presenting a comprehensive comparison of all remaining source parameters.

\begin{figure*}
\centering
\begin{subfigure}[b]{0.88\linewidth}
\centering
\includegraphics[width=\textwidth,height=12cm]{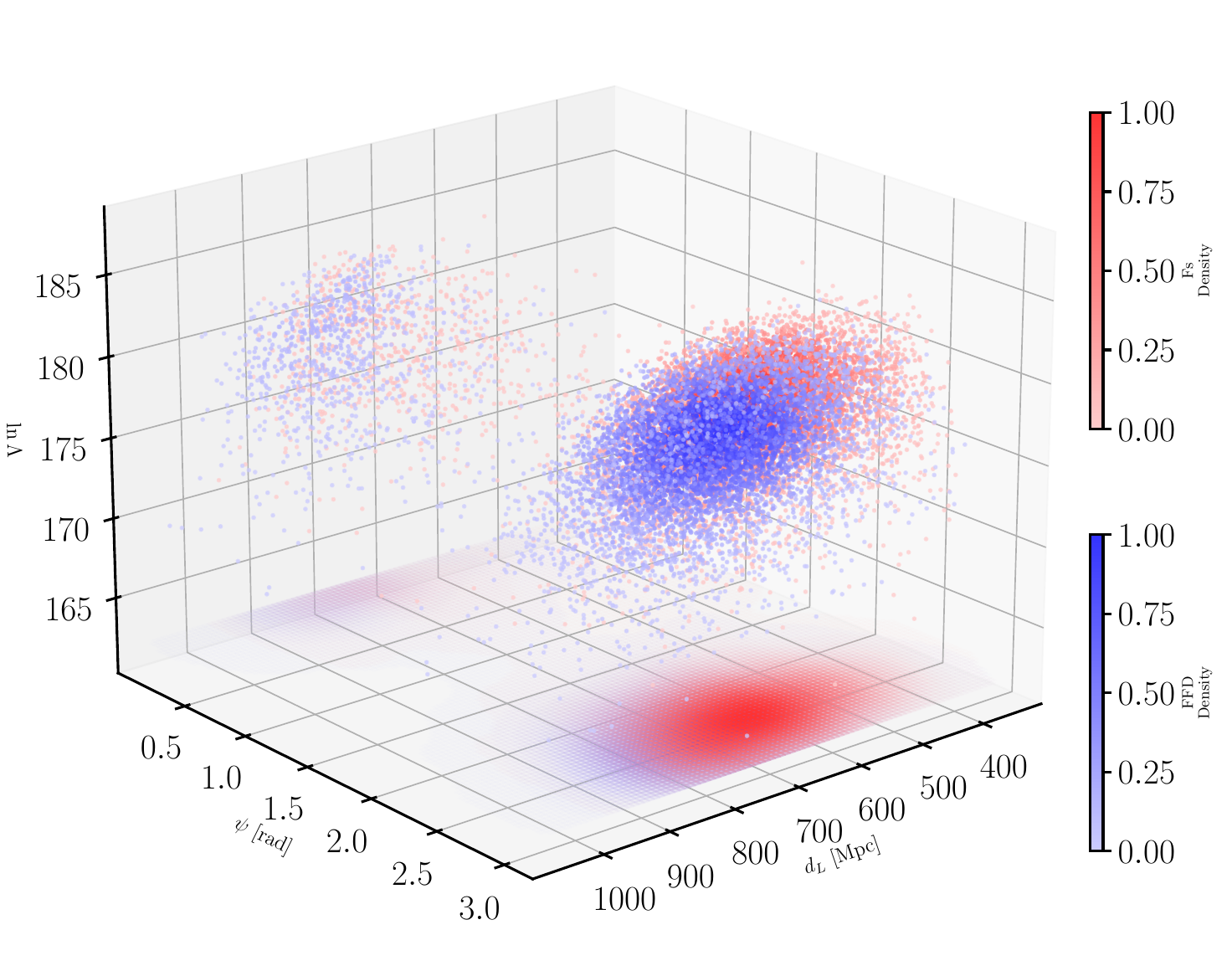}
\end{subfigure}\\
\begin{subfigure}[b]{0.48\linewidth}
\centering
\includegraphics[width=\textwidth,height=8cm]{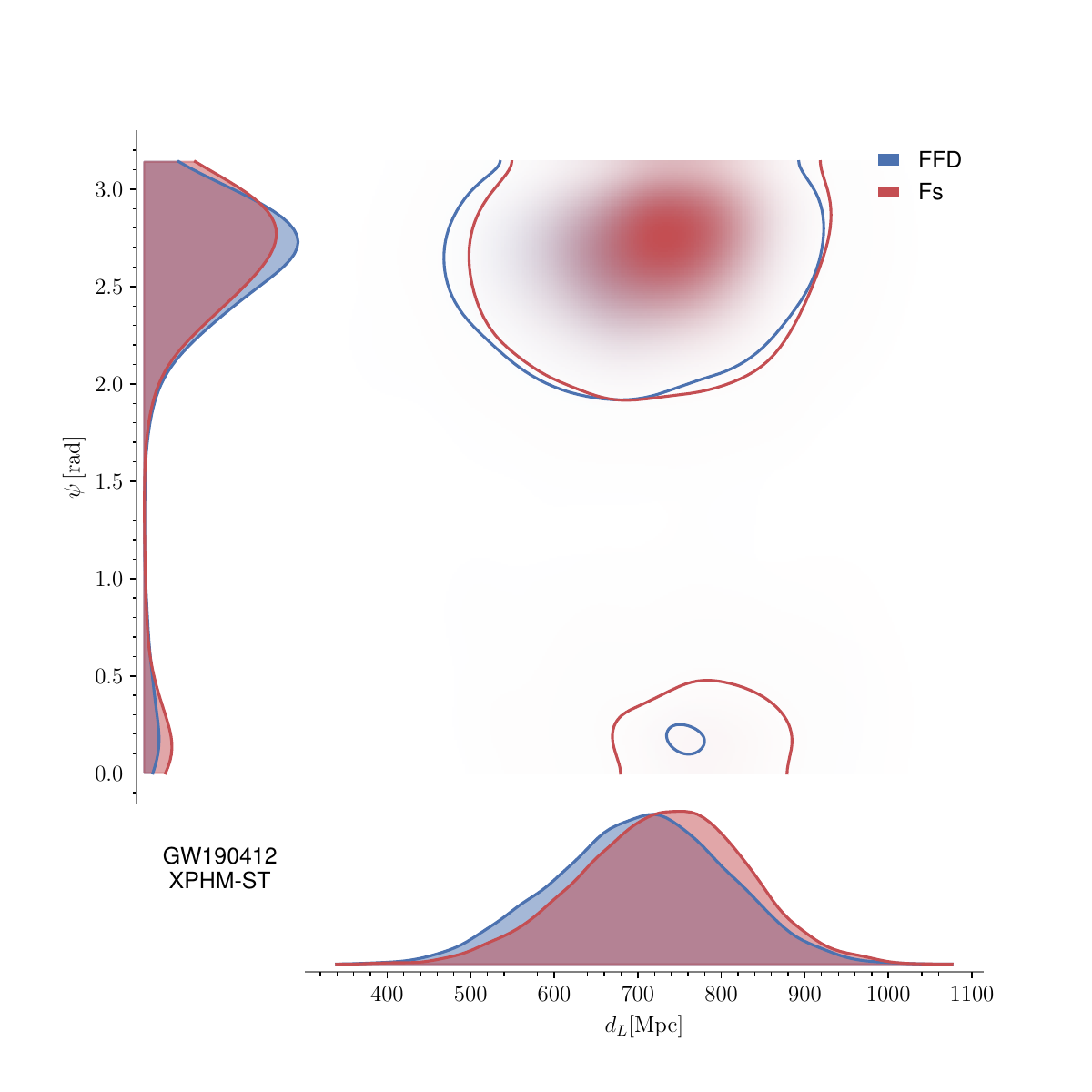}
\end{subfigure}%
\begin{subfigure}[b]{0.48\linewidth}
\centering
\includegraphics[width=\textwidth,height=8cm]{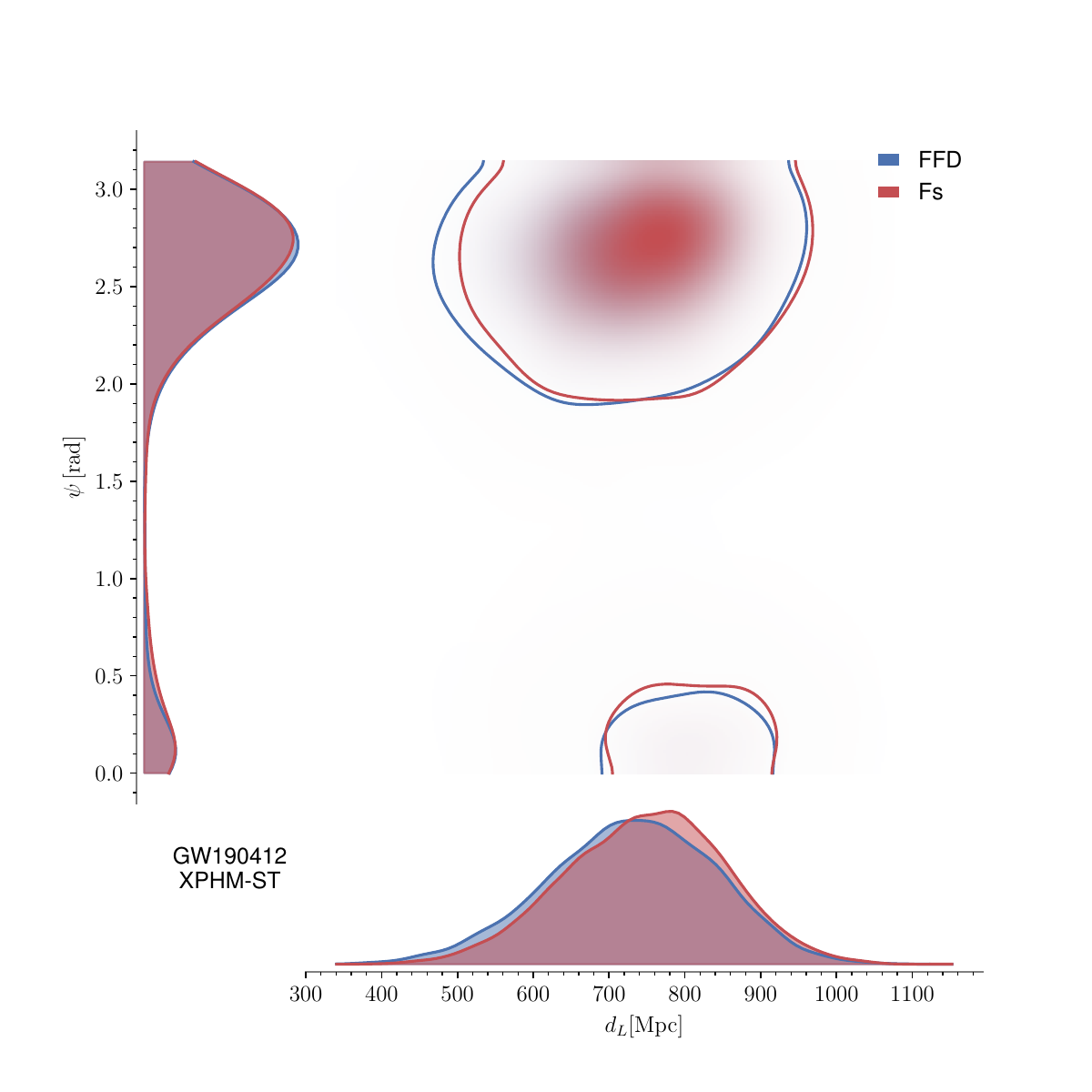}
\end{subfigure}%
\caption{
Comparison of the luminosity distance $d_L$ and polarization angle $\psi$ inferred for GW190412 using the $\fs$ (Fs) and \ac{FFD} likelihood approaches.
The top panel shows a three-dimensional scatter plot of samples in the $(d_L, \psi, \ln \Lambda)$ space obtained with both methods when calibration uncertainties are included. 
Each point corresponds to a posterior sample, with the color indicating the relative sample density. 
This representation illustrates that the high-likelihood regions identified by the two approaches occupy consistent regions of parameter space.
The lower left panel displays the joint posterior distributions of $(d_L, \psi)$ obtained without accounting for calibration errors, while the lower right panel shows the corresponding posteriors when calibration uncertainties are included.
Contours indicate the marginalized two-dimensional credible regions at $90\%$ level, and the one-dimensional projections are shown along the axes.
When calibration errors are neglected, small but noticeable differences between the $\fs$ and FFD posteriors are present, particularly along the degeneracy direction between $d_L$ and $\psi$.
Including calibration uncertainties broadens the posteriors and leads to substantially improved agreement between the two methods, indicating that calibration marginalization plays an important role in harmonizing likelihood-based inference schemes for these linear extrinsic parameters.
}
\label{fig:dl-psi1}
\end{figure*}

\subsubsection{Reconstruction of Analytically Maximized Parameters}
The luminosity distance, $d_L$, and polarization angle, $\psi$, are treated fundamentally differently in the two frameworks. In the \ac{FFD} approach, they are sampled directly along with all other parameters. In the $\fs$ method, they are analytically maximized, and their posteriors must be reconstructed in a post-processing step, as detailed in Sec.~\ref{ssec:plp}. 
Fig~\ref{fig:dl-psi1} presents a detailed comparison of the luminosity distance and polarization angle inferred using the $\fs$ and FFD approaches.
These two parameters enter the detector response linearly and are therefore particularly sensitive to differences in likelihood construction and marginalization strategies.
GW190412, with its asymmetric mass ratio and measurable higher-order mode content, provides sufficient information to partially break the distance--inclination--polarization degeneracies, making it an ideal system for such a comparison.

\begin{table*}[hptb]
\begin{ruledtabular}
\begin{tabular}{l | c c c | c c c}
 & \multicolumn{3}{c|}{No Calibration Error (noc)} & \multicolumn{3}{c}{Calibration Error (c)} \\
 Parameter & Fs-noc & FFD-noc & JSD & Fs-c & FFD-c & JSD \\
\hline
$t_c-t_0 [\mathrm{ms}]$ & $0.96^{+0.93}_{-0.92}$ & $1.01^{+0.98}_{-0.87}$ & 0.004 & $0.96^{+0.91}_{-0.83}$ & $0.95^{+0.89}_{-0.84}$ & 0.003 \\
$\mathcal{M} [M_{\odot}]$ & $15.25^{+0.30}_{-0.21}$ & $15.26^{+0.31}_{-0.20}$ & 0.003 & $15.25^{+0.33}_{-0.21}$ & $15.27^{+0.36}_{-0.21}$ & 0.005 \\
$q$ & $0.28^{+0.09}_{-0.07}$ & $0.28^{+0.08}_{-0.07}$ & 0.003 & $0.28^{+0.10}_{-0.07}$ & $0.27^{+0.09}_{-0.07}$ & 0.012 \\
$a_{1}$ & $0.37^{+0.16}_{-0.17}$ & $0.38^{+0.16}_{-0.16}$ & 0.004 & $0.36^{+0.16}_{-0.19}$ & $0.39^{+0.16}_{-0.17}$ & 0.017 \\
$a_{2}$ & $0.54^{+0.41}_{-0.48}$ & $0.54^{+0.39}_{-0.47}$ & 0.006 & $0.53^{+0.41}_{-0.47}$ & $0.51^{+0.44}_{-0.45}$ & 0.004 \\
$\theta_{1} [\mathrm{rad}]$ & $0.59^{+0.42}_{-0.35}$ & $0.63^{+0.41}_{-0.34}$ & 0.009 & $0.57^{+0.47}_{-0.37}$ & $0.60^{+0.41}_{-0.36}$ & 0.007 \\
$\theta_{2} [\mathrm{rad}]$ & $1.33^{+1.21}_{-0.96}$ & $1.30^{+1.15}_{-0.90}$ & 0.004 & $1.32^{+1.21}_{-0.95}$ & $1.34^{+1.24}_{-0.95}$ & 0.002 \\
$\phi_{12} [\mathrm{rad}]$ & $2.55^{+3.41}_{-2.28}$ & $2.65^{+3.21}_{-2.28}$ & 0.005 & $2.69^{+3.29}_{-2.39}$ & $2.79^{+3.18}_{-2.48}$ & 0.003 \\
$\phi_{JL} [\mathrm{rad}]$ & $1.00^{+5.14}_{-0.89}$ & $1.02^{+5.09}_{-0.88}$ & 0.005 & $1.16^{+5.01}_{-1.04}$ & $1.12^{+5.07}_{-1.01}$ & 0.010 \\
$\theta_{JN} [\mathrm{rad}]$ & $0.85^{+1.37}_{-0.29}$ & $0.79^{+1.54}_{-0.27}$ & 0.027 & $0.85^{+1.29}_{-0.31}$ & $0.76^{+0.39}_{-0.28}$ & 0.044 \\
$\phi [\mathrm{rad}]$ & $2.00^{+1.15}_{-0.69}$ & $2.06^{+1.22}_{-0.69}$ & 0.005 & $2.04^{+1.30}_{-0.71}$ & $2.04^{+1.35}_{-0.75}$ & 0.002 \\
$\alpha [\mathrm{rad}]$ & $3.81^{+0.04}_{-0.06}$ & $3.81^{+0.04}_{-0.08}$ & 0.003 & $3.81^{+0.04}_{-0.06}$ & $3.81^{+0.03}_{-0.05}$ & 0.005 \\
$\delta [\mathrm{rad}]$ & $0.63^{+0.03}_{-0.04}$ & $0.63^{+0.03}_{-0.05}$ & 0.004 & $0.63^{+0.03}_{-0.04}$ & $0.63^{+0.03}_{-0.04}$ & 0.003 \\
$d_{L} [\mathrm{Mpc}]$ & $732.2^{+150.9}_{-172.5}$ & $706.2^{+159.4}_{-175.7}$ & 0.014 & $751.2^{+161.9}_{-179.1}$ & $732.2^{+171.2}_{-192.8}$ & 0.008 \\
$\psi [\mathrm{rad}]$ & $2.66^{+0.42}_{-2.54}$ & $2.65^{+0.38}_{-2.24}$ & 0.017 & $2.64^{+0.42}_{-2.53}$ & $2.63^{+0.42}_{-2.53}$ & 0.007 \\
\hline
$N_s$ & $7016$ & $7559$ & - & $7140$ & $7141$ & - \\
$T_s [h]$ & $5.5$ & $11.7$ & - & $4.5$ & $20.1$ & - \\
$\ln\mathcal{Z}$ & $146.7$ & $145.9$ & - & $146.7$ & $146.6$ & - \\
\end{tabular}
\caption{Posterior constraints for GW190412 estimated by $\fs$ (Fs) and FFD methods under two scenarios: without calibration error (noc) and with calibration error (c). 
For each parameter, we report the median and symmetric $90\%$ credible intervals.
The \ac{JSD} quantifies the similarity between the posterior distributions inferred by the two methods under the same calibration assumptions.
The lower rows list the number of posterior samples ($N_s$), total sampling time ($T_s$), and the log Bayesian evidence ($\ln\mathcal{Z}$) for each analysis.
}
\label{tab:GW190412_allp}
\end{ruledtabular}
\begin{tablenotes}
    \small
    \item[a] 1. Parameter definitions follow the conventions used in the {\sc PESummary} package~\cite{Hoy:2020vys}.
    \item[b] 2. The geocentric time $t_c$ is reported as an offset from the GW190412 trigger time, $t_0=1239082262.18$ s.
\end{tablenotes}
\end{table*}

In the absence of calibration uncertainties, the two methods yield broadly consistent but not identical joint posteriors in the $(d_L,\psi)$ plane.
The residual discrepancies are most pronounced along the elongated degeneracy direction, reflecting the different treatments of the linear extrinsic parameters in the two inference schemes.
When calibration errors are incorporated, the posterior distributions widen slightly, as expected, as these uncertainties are known to be partially degenerate with the luminosity distance and inclination \citep{Vitale:2011wu,Payne:2020myg,Sun:2020wke}. However, the agreement between the $\fs$ and FFD results improves markedly.

These results demonstrate that calibration marginalization mitigates small method-dependent differences in the inference of linear extrinsic parameters, and leads to highly consistent posteriors between accelerated likelihood approaches.
In particular, for $d_L$ and $\psi$, the inclusion of calibration uncertainties effectively absorbs residual systematic differences between the $\fs$ and FFD formulations. 
This improved agreement is quantitatively reflected in Table~\ref{tab:GW190412_allp}. 
Specifically, upon introducing calibration uncertainty, the \ac{JSD} for the luminosity distance decreases from $0.014$ to $0.008$, while the \ac{JSD} for the polarization angle drops from $0.017$ to $0.007$.
These reductions confirm that including calibration uncertainty yields nearly indistinguishable posterior constraints for these parameters.

\subsubsection{Comparison of All Sampled Source Parameters}
Table~\ref{tab:GW190412_allp} summarizes the full set of posterior constraints obtained for GW190412 using the $\fs$ and FFD approaches under identical analysis configurations.
Across both calibration scenarios, we find excellent agreement between the two methods for all intrinsic and extrinsic parameters.
This is quantitatively confirmed by the \ac{JSD} values, which remain at the level of $\mathcal{O}(10^{-2})$ or below for the majority of parameters, indicating nearly indistinguishable posterior distributions.

Intrinsic parameters such as the chirp mass, mass ratio, and component spins show particularly strong consistency between the two methods, independent of whether calibration uncertainties are included.
Extrinsic parameters, including sky location, coalescence phase, and polarization angle, also exhibit close agreement, with slightly larger but still small JSD values reflecting their broader and more degenerate posteriors.
As already highlighted in the $(d_L,\psi)$ comparison, the inclusion of calibration uncertainties systematically improves the consistency between the two approaches for parameters entering the detector response linearly.

\begin{figure*}
\centering
\begin{subfigure}[b]{0.48\linewidth}
\centering
\includegraphics[width=\textwidth,height=6.6cm]{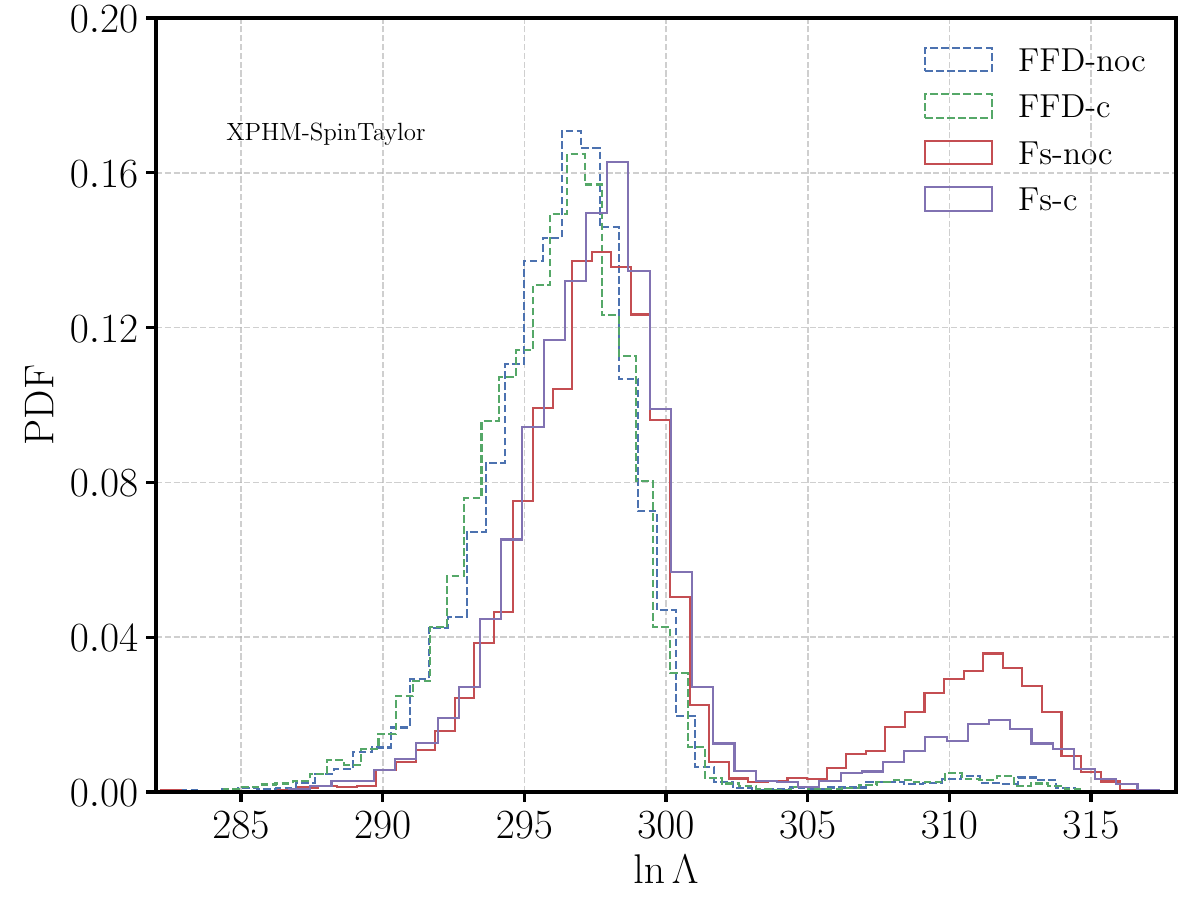}
\end{subfigure}%
\begin{subfigure}[b]{0.48\linewidth}
\centering
\includegraphics[width=\textwidth,height=6.6cm]{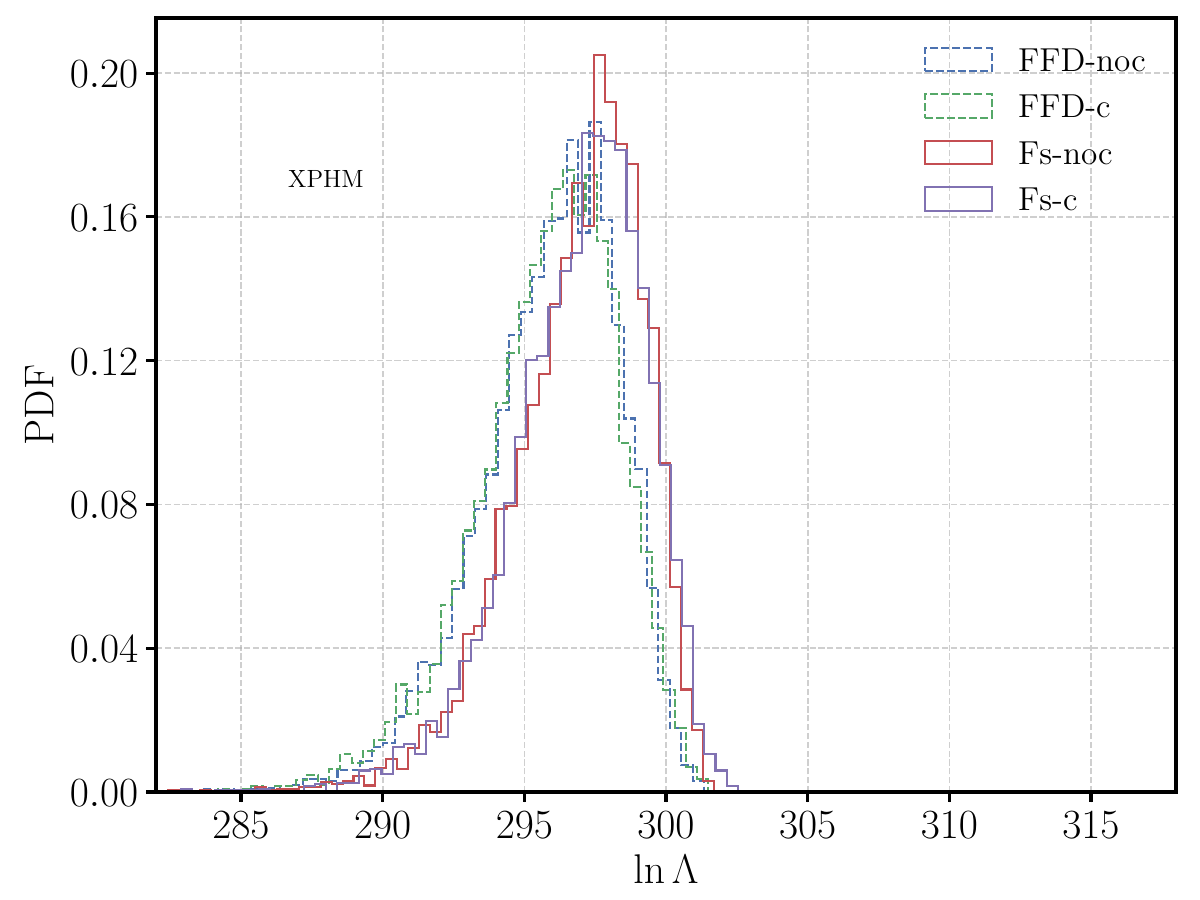}
\end{subfigure}%
\caption{
Comparison of the log-likelihood distributions for GW190814 obtained using the $\fs$ (Fs) and FFD inference methods under different waveform and calibration assumptions.
The left panel shows results obtained with the \texttt{IMRPhenomXPHM-SpinTaylor} waveform, while the right panel corresponds to the \texttt{IMRPhenomXPHM} waveform.
For each waveform, we present four analyses: Fs and FFD, each performed with (c) and without (noc) including calibration uncertainties.
For the SpinTaylor variant, the likelihood distribution exhibits a pronounced bimodal structure, whereas the \texttt{IMRPhenomXPHM} analysis yields a single dominant mode consistent with the LVK results.
Including calibration uncertainty systematically suppresses the secondary likelihood peak in the Fs analysis, reducing the relative weight of the subdominant mode.
}
\label{fig:GW190814bbh_lls}
\end{figure*}

In terms of computational performance, the $\fs$ method consistently yields a higher Bayesian evidence, as shown in Table~\ref{tab:GW190412_allp}, demonstrating that the accelerated inference scheme does not compromise statistical fidelity.
However, the sampling efficiency differs drastically between the two approaches. 
As detailed in Table~\ref{tab:GW190412_allp}, in the absence of calibration errors, the $\fs$ method reduces the total sampling time ($T_s$) by approximately $53\%$ ($5.5$ h vs $11.7$ h). 
Notably, this computational advantage becomes significantly more pronounced when calibration uncertainties are included. 
While the computational cost of the FFD analysis nearly doubles to $20.1$ h in this scenario, the $\fs$ analysis completes in just $4.5$ h—a speed-up factor of approximately $4.5$.
These results confirm that the $\fs$ method provides a robust and computationally efficient framework for complex \ac{GW} parameter estimation, offering substantial time savings that become even more critical as the complexity of the inference problem increases.

Overall, the results presented here establish GW190412 as a stringent validation case for accelerated likelihood-based inference, and demonstrate that both the $\fs$ and FFD approaches can reliably recover high-dimensional posterior distributions even in the presence of higher-order modes, precession, and calibration uncertainties.

\subsection{Analysis of GW190814: A High-Mass-Ratio System}
Following the comparative study of GW190412, we extend our analysis to GW190814, a signal detected by the LVK detectors during the third observing run. GW190814 is a particularly distinct event characterized by the most unequal mass ratio observed to date ($q\approx 0.112$) and a secondary component mass of approximately $2.6\Msun$, which places it in the lower mass gap between known neutron stars and black holes. The highly asymmetric masses amplify the emission of higher-order multipoles, making their contribution to the signal significant. Furthermore, the unique parameter space of this event provides a stringent testing ground for waveform systematics and inference methods, particularly regarding the interplay between spin-precession effects and higher-order modes.
To evaluate the performance of the FFD method and the $\fs$ method on this source, we perform four distinct analyses. These consist of applying both methods under two conditions: one neglecting calibration errors and one marginalizing over calibration uncertainties. 

\begin{figure*}
\centering
\begin{subfigure}[b]{0.88\linewidth}
\centering
\includegraphics[width=\textwidth,height=12cm]{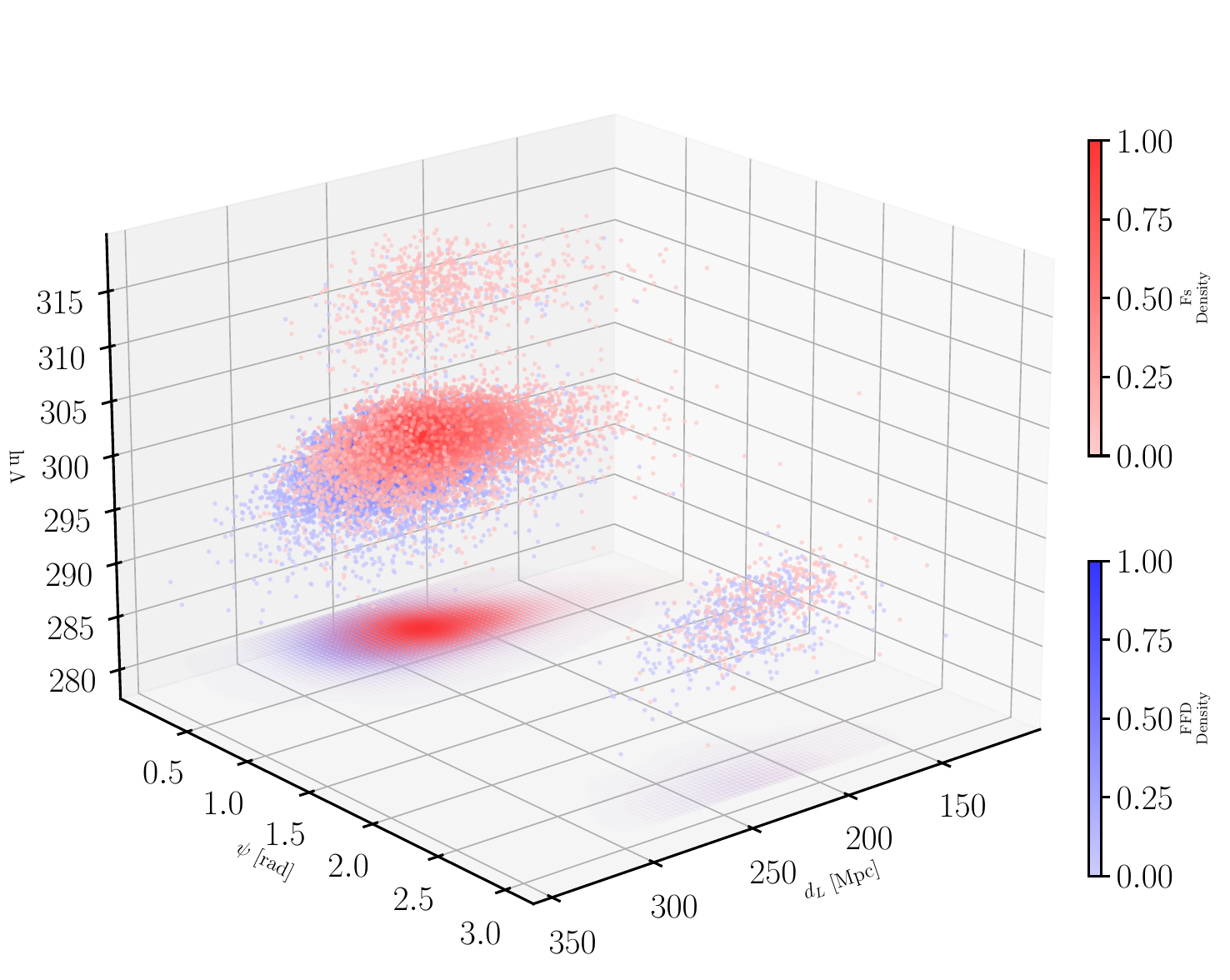}
\end{subfigure}\\
\begin{subfigure}[b]{0.48\linewidth}
\centering
\includegraphics[width=\textwidth,height=9cm]{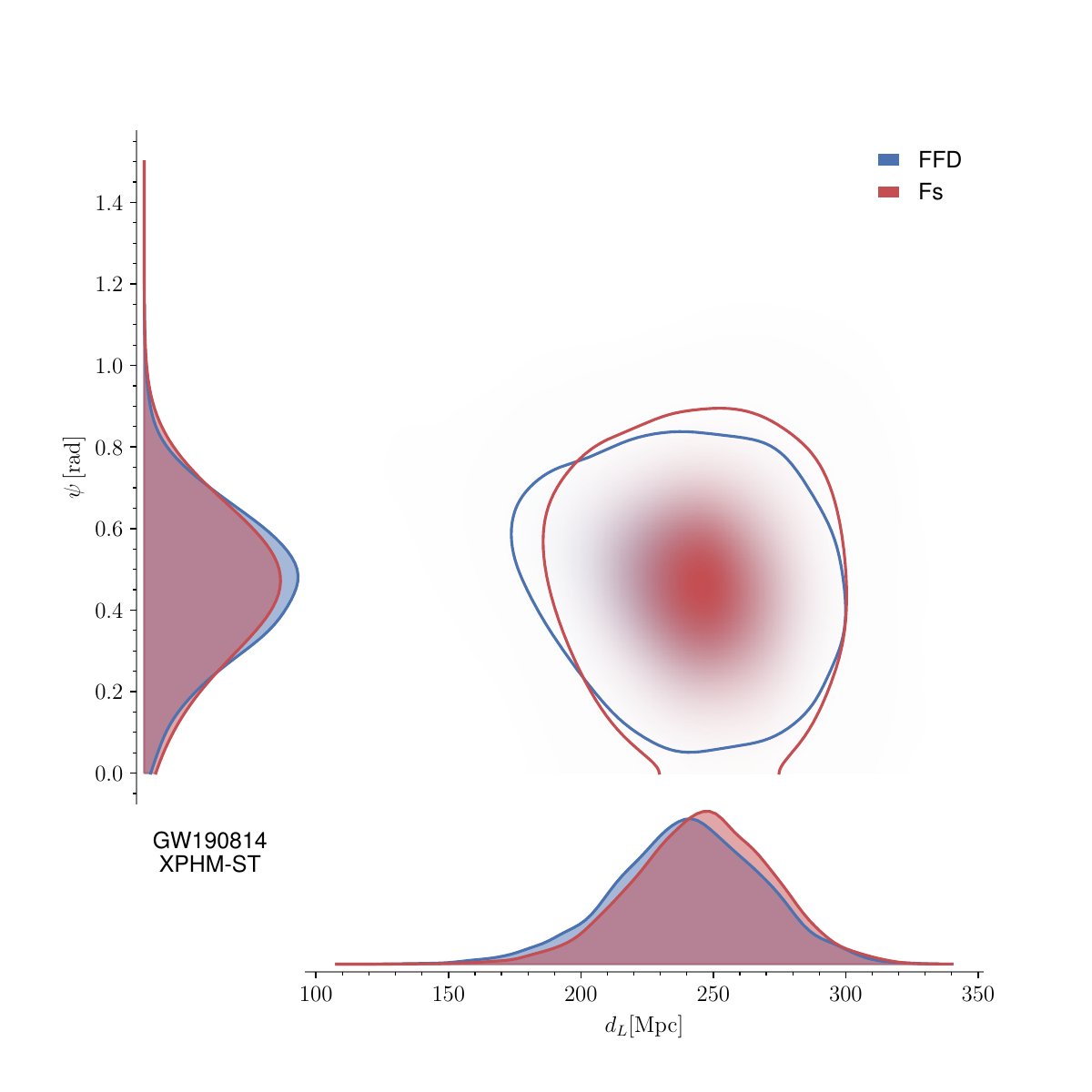}
\end{subfigure}%
\begin{subfigure}[b]{0.48\linewidth}
\centering
\includegraphics[width=\textwidth,height=9cm]{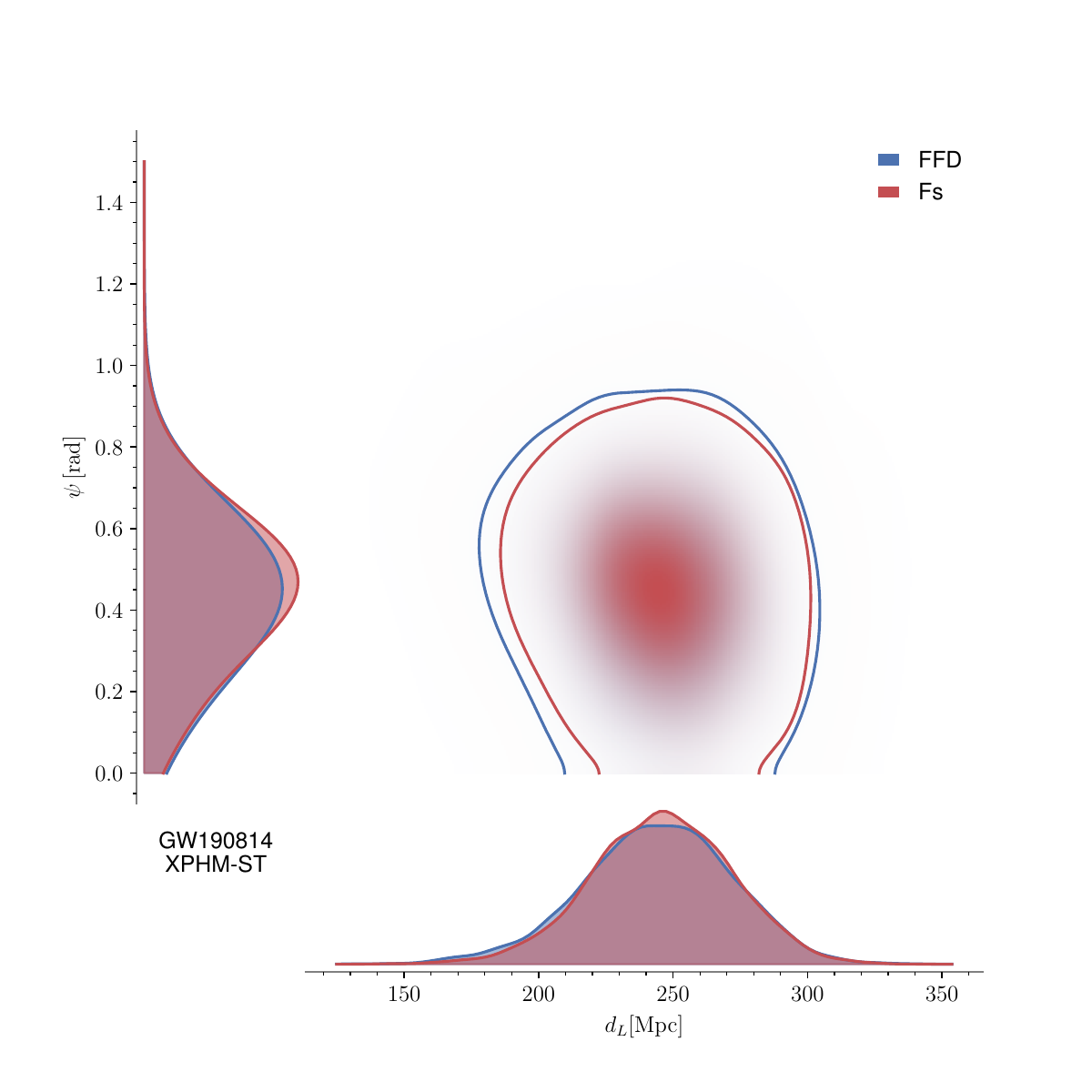}
\end{subfigure}%
\caption{
Joint distributions of luminosity distance $d_L$ and polarization angle $\psi$ for GW190814 inferred using the \texttt{IMRPhenomXPHM-SpinTaylor} waveform.
The arrangement of panels is the same as in Fig.~\ref{fig:dl-psi1}.
While both methods recover consistent primary modes, secondary structures associated with the bimodal likelihood are more visible in the $\fs$ cases.
Marginalization over calibration uncertainties reduces the prominence of these secondary features and leads to improved agreement between the $\fs$ and FFD posteriors.
}
\label{fig:dl-psi2}
\end{figure*}

Regarding the choice of waveform approximants, our approach for GW190814 deviates slightly from the previous section. While the GW190412 analysis relied exclusively on the \texttt{IMRPhenomXPHM-SpinTaylor} model, our preliminary analysis of GW190814 using this waveform revealed a bimodal structure in the likelihood distribution, as shown in Fig.~\ref{fig:GW190814bbh_lls}. In contrast, the posterior distributions reported by the LVK collaboration—based on \texttt{IMRPhenomXPHM}—exhibit a single peak. To investigate this discrepancy and ensure robust parameter estimation, we additionally employ the \texttt{IMRPhenomXPHM} approximant. We find that \texttt{IMRPhenomXPHM} yields a unimodal likelihood distribution consistent with LVK results, allowing for a direct comparison with the SpinTaylor variant. Furthermore, given the ambiguous nature of the secondary object—which may be either a heavy neutron star or a light black hole—we also conduct a separate analysis using the \texttt{IMRPhenomNSBH} waveform model to assess potential tidal signatures.

\begin{table*}[hptb]
\begin{ruledtabular}
\begin{tabular}{l | c c c | c c c}
 & \multicolumn{3}{c|}{No Calibration Error (noc)} & \multicolumn{3}{c}{Calibration Error (c)} \\
 Parameter & Fs-noc & FFD-noc & JSD & Fs-c & FFD-c & JSD \\
\hline
$t_c-t_0 [\mathrm{ms}]$ & $-0.09^{+0.33}_{-0.33}$ & $-0.11^{+0.33}_{-0.33}$ & 0.007 & $-0.10^{+0.35}_{-0.34}$ & $-0.09^{+0.38}_{-0.35}$ & 0.003 \\
$\mathcal{M} [M_{\odot}]$ & $6.41^{+0.02}_{-0.04}$ & $6.41^{+0.02}_{-0.04}$ & 0.005 & $6.41^{+0.02}_{-0.05}$ & $6.41^{+0.02}_{-0.04}$ & 0.003 \\
$q$ & $0.11^{+0.01}_{-0.01}$ & $0.11^{+0.01}_{-0.01}$ & 0.025 & $0.11^{+0.01}_{-0.01}$ & $0.11^{+0.01}_{-0.01}$ & 0.008 \\
$a_{1}$ & $0.03^{+0.08}_{-0.02}$ & $0.04^{+0.07}_{-0.04}$ & 0.034 & $0.03^{+0.09}_{-0.03}$ & $0.04^{+0.08}_{-0.04}$ & 0.012 \\
$a_{2}$ & $0.51^{+0.42}_{-0.45}$ & $0.49^{+0.43}_{-0.42}$ & 0.004 & $0.50^{+0.44}_{-0.44}$ & $0.48^{+0.45}_{-0.43}$ & 0.003 \\
$\theta_{1} [\mathrm{rad}]$ & $1.55^{+1.17}_{-1.15}$ & $1.60^{+1.09}_{-1.12}$ & 0.005 & $1.51^{+1.21}_{-1.15}$ & $1.51^{+1.20}_{-1.14}$ & 0.004 \\
$\theta_{2} [\mathrm{rad}]$ & $1.61^{+0.94}_{-1.11}$ & $1.51^{+1.01}_{-1.03}$ & 0.009 & $1.59^{+1.03}_{-1.10}$ & $1.56^{+1.06}_{-1.08}$ & 0.004 \\
$\phi_{12} [\mathrm{rad}]$ & $3.50^{+2.35}_{-3.05}$ & $3.20^{+2.65}_{-2.77}$ & 0.045 & $3.47^{+2.41}_{-3.06}$ & $3.15^{+2.73}_{-2.80}$ & 0.016 \\
$\phi_{JL} [\mathrm{rad}]$ & $2.03^{+3.89}_{-1.76}$ & $1.59^{+4.13}_{-1.30}$ & 0.018 & $2.08^{+3.89}_{-1.77}$ & $1.86^{+4.01}_{-1.58}$ & 0.006 \\
$\theta_{JN} [\mathrm{rad}]$ & $0.90^{+1.37}_{-0.24}$ & $0.85^{+0.79}_{-0.22}$ & 0.024 & $0.90^{+1.40}_{-0.24}$ & $0.85^{+1.54}_{-0.24}$ & 0.035 \\
$\phi [\mathrm{rad}]$ & $2.75^{+0.51}_{-1.54}$ & $2.74^{+0.51}_{-1.50}$ & 0.003 & $2.76^{+0.51}_{-1.60}$ & $2.75^{+0.53}_{-1.49}$ & 0.005 \\
$\alpha [\mathrm{rad}]$ & $0.22^{+0.16}_{-0.03}$ & $0.22^{+0.08}_{-0.03}$ & 0.009 & $0.23^{+0.17}_{-0.03}$ & $0.23^{+0.17}_{-0.03}$ & 0.006 \\
$\delta [\mathrm{rad}]$ & $-0.44^{+0.03}_{-0.12}$ & $-0.44^{+0.04}_{-0.07}$ & 0.006 & $-0.44^{+0.04}_{-0.12}$ & $-0.44^{+0.04}_{-0.13}$ & 0.007 \\
$d_{L} [\mathrm{Mpc}]$ & $246.3^{+40.9}_{-44.3}$ & $241.1^{+43.5}_{-48.5}$ & 0.009 & $246.0^{+41.8}_{-43.6}$ & $244.7^{+43.5}_{-48.1}$ & 0.005 \\
$\psi [\mathrm{rad}]$ & $0.47^{+0.37}_{-0.32}$ & $0.47^{+0.28}_{-0.30}$ & 0.011 & $0.46^{+0.50}_{-0.34}$ & $0.46^{+2.54}_{-0.35}$ & 0.004 \\
\hline
$N_s$ & $7484$ & $6348$ & - & $6817$ & $6780$ & - \\
$T_s [h]$ & $34.3$ & $44.1$ & - & $59.6$ & $95.6$ & - \\
$\ln\mathcal{Z}$ & $261.3$ & $261.0$ & - & $261.4$ & $260.1$ & - \\
\end{tabular}
\caption{
Posterior parameter estimates for GW190814 obtained using the \texttt{IMRPhenomXPHM-SpinTaylor} waveform with the $\fs$ (Fs) and FFD methods.
Results are shown for analyses performed without (noc) and with (c) marginalization over calibration uncertainties.
For each parameter, we report the median and symmetric $90\%$ credible intervals, together with the \ac{JSD} quantifying the agreement between the $\fs$ and FFD posterior distributions.
Also listed are the number of posterior samples ($N_s$), total sampling time ($T_s$), and the log Bayesian evidence ($\ln\mathcal{Z}$).
Including calibration uncertainty systematically reduces the JSD for most parameters, indicating improved consistency between the two inference methods.
}
\label{tab:GW190814xphmST_allp}
\end{ruledtabular}
\begin{tablenotes}
    \small
    \item[a] 1. The geocentric time $t_c$ is reported as an offset from the GW190814 trigger time, $t_0=1249852256.99$ s.
\end{tablenotes}
\end{table*}

The analysis of this event is structured as follows: First, we compare the likelihood distributions obtained from \texttt{IMRPhenomXPHM} and \texttt{IMRPhenomXPHM-SpinTaylor} to address the observed bimodality. Second, we examine the joint posterior distributions of luminosity distance and polarization angle. Third, we present a comprehensive comparison of the constraints on intrinsic and extrinsic parameters. Finally, we report the results obtained using the \texttt{IMRPhenomNSBH} template.

\subsubsection{Results of IMRPhenomXPHM-SpinTaylor}
When analyzed with the \texttt{IMRPhenomXPHM-SpinTaylor} waveform, the log-likelihood distribution exhibits a clear bimodal structure for both the $\fs$ and FFD runs (Fig.~\ref{fig:GW190814bbh_lls}).
Inspecting the high-dimensional likelihood landscape suggests that the subdominant mode, which occurs at higher likelihood values, is associated with a different region of extrinsic-parameter space, with luminosity distance and polarization angle being among the most affected directions (top panel of Fig.~\ref{fig:dl-psi2}).
Notably, the $\fs$ method appears to explore this subdominant, higher-likelihood mode more extensively. However, this bimodality observed in the distribution of $\ln\Lambda$ does not translate into a distinctly bimodal marginalized posterior in the $(d_L,\psi)$ plane.

\begin{figure*}
\centering
\begin{subfigure}[b]{0.88\linewidth}
\centering
\includegraphics[width=\textwidth,height=12cm]{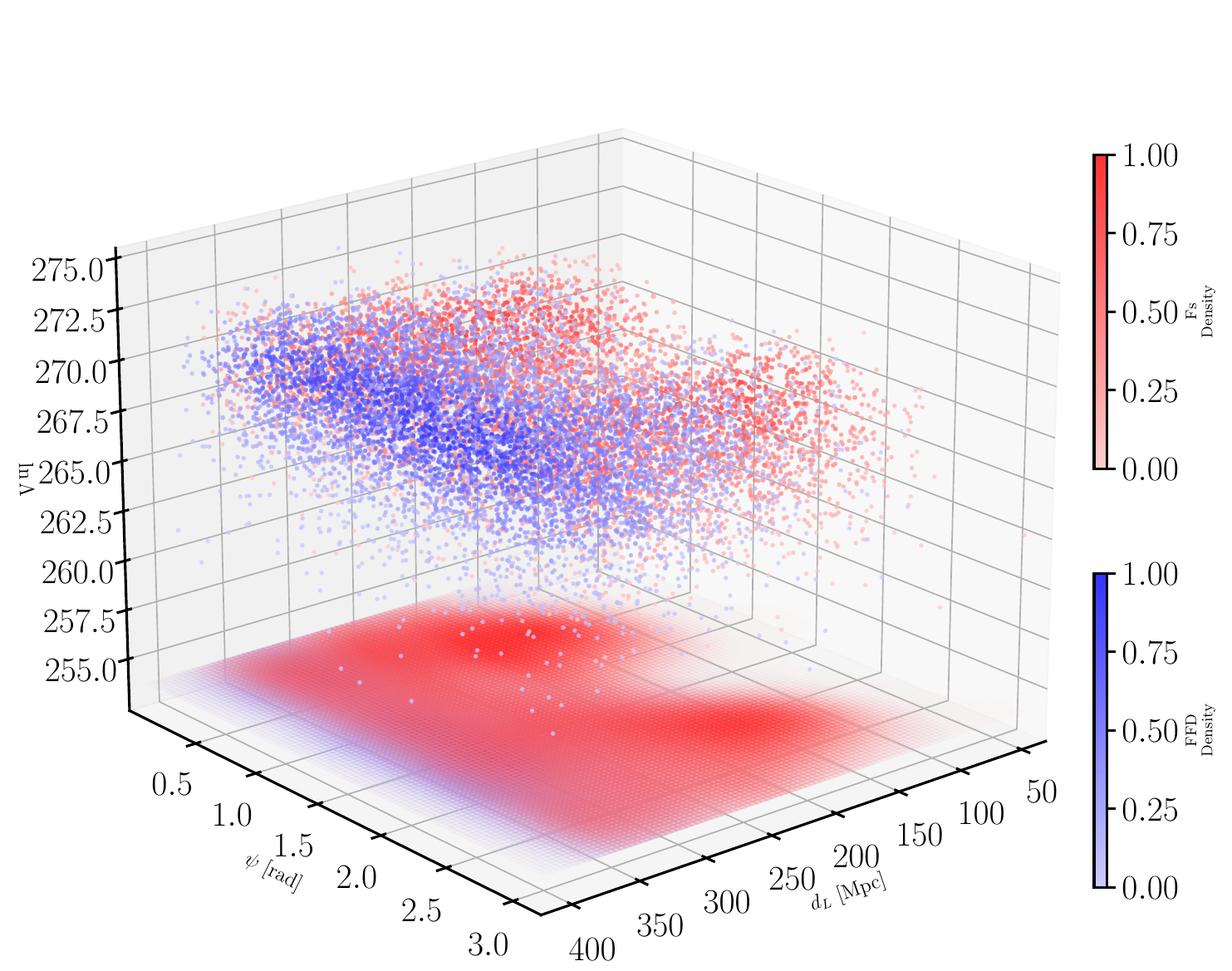}
\end{subfigure}\\
\begin{subfigure}[b]{0.48\linewidth}
\centering
\includegraphics[width=\textwidth,height=9cm]{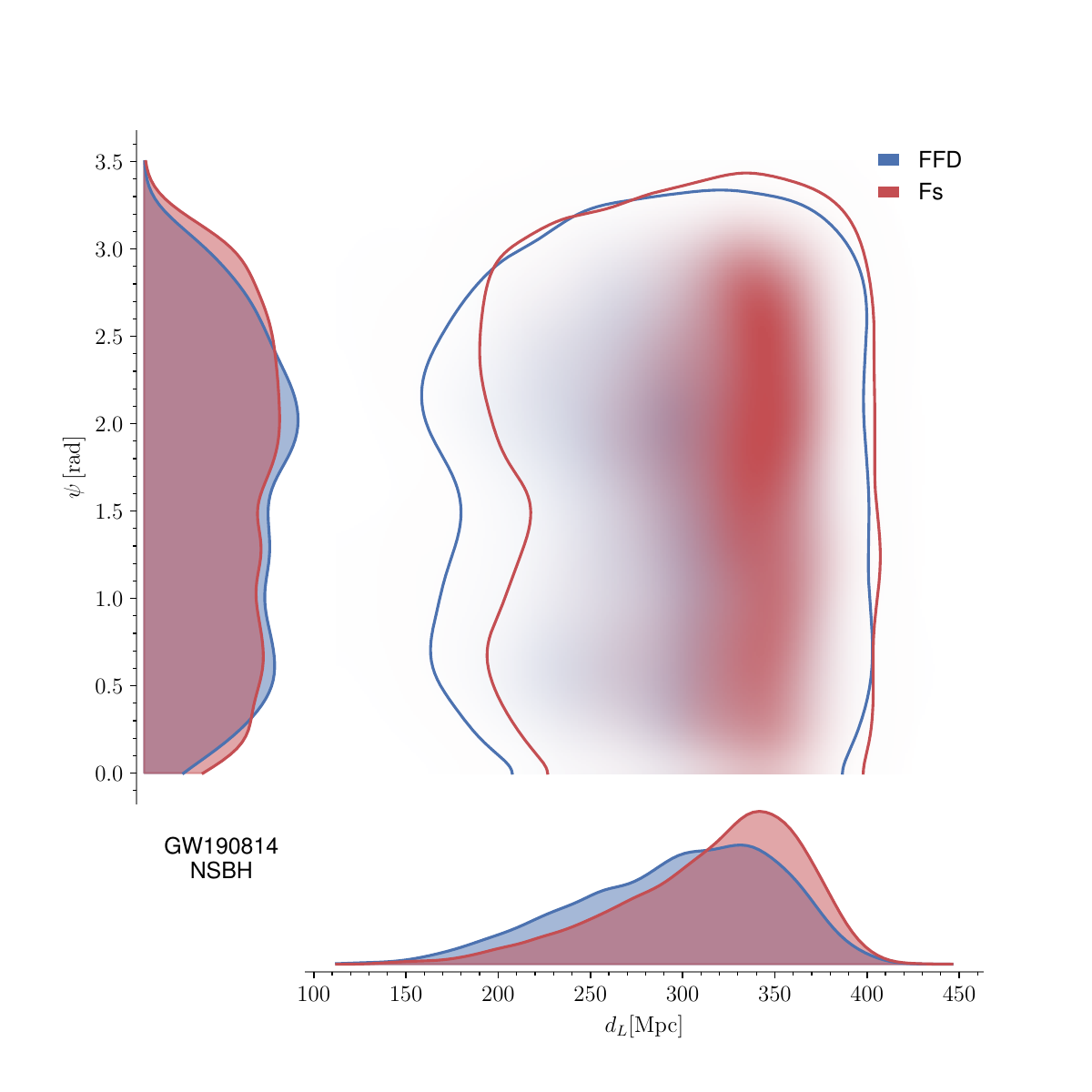}
\end{subfigure}%
\begin{subfigure}[b]{0.48\linewidth}
\centering
\includegraphics[width=\textwidth,height=9cm]{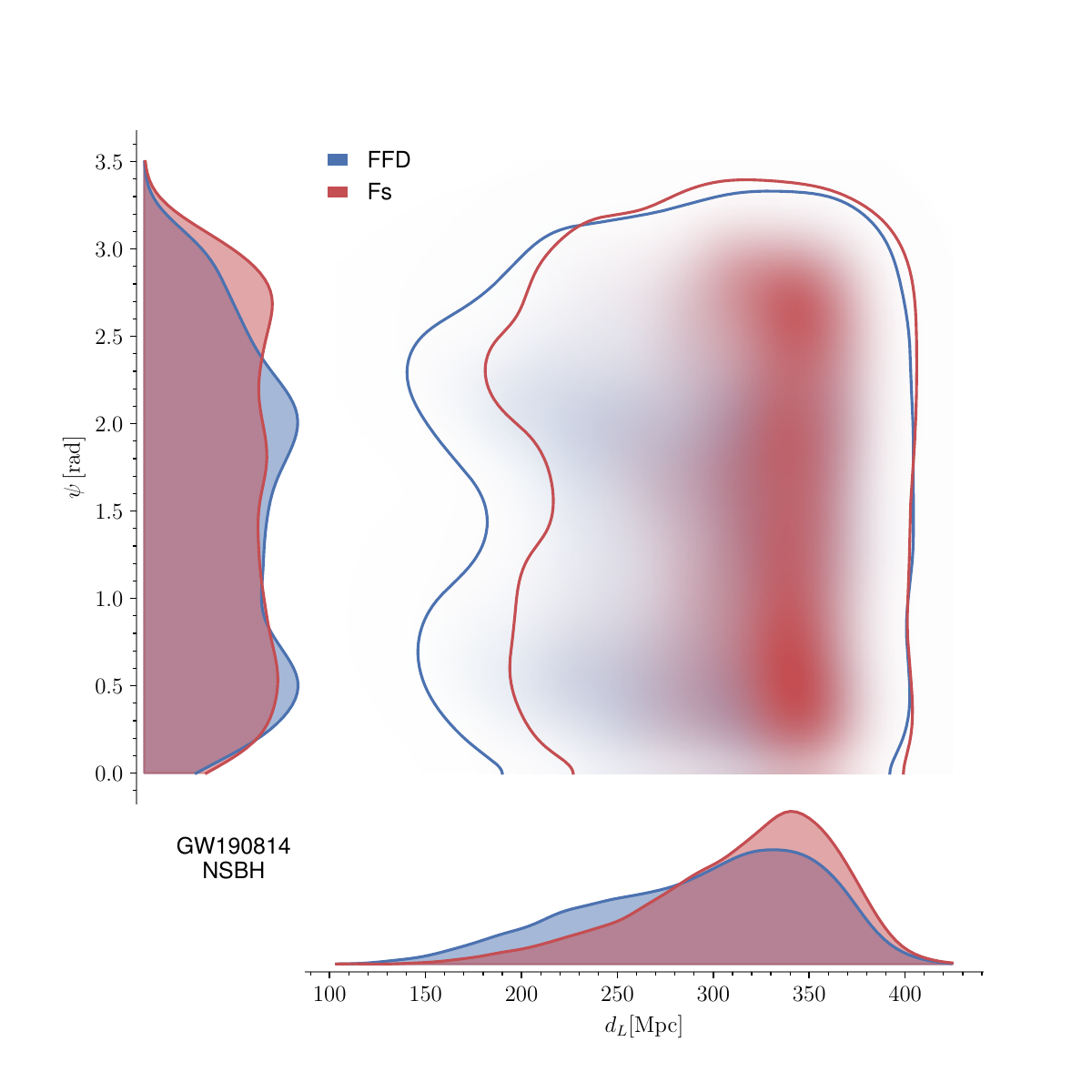}
\end{subfigure}%
\caption{
Same as Fig.~\ref{fig:dl-psi1}, but for the analysis of GW190814 under the NSBH hypothesis using the \texttt{IMRPhenomNSBH} waveform model.
}
\label{fig:dl-psi3}
\end{figure*}

Such bimodality is absent in the LVK posterior results, which were obtained using the \texttt{IMRPhenomXPHM} waveform.
To investigate this discrepancy, we repeat the analysis using \texttt{IMRPhenomXPHM}.
In this case, the likelihood distribution becomes unimodal and closely matches the LVK results, indicating that the secondary likelihood peak observed with the SpinTaylor variant is waveform-dependent rather than a sampling artifact.
This behavior highlights the sensitivity of GW190814 to subtle modeling differences in the treatment of spin precession and higher-order modes.

\begin{table*}[hptb]
\begin{ruledtabular}
\begin{tabular}{l | c c c | c c c}
 & \multicolumn{3}{c|}{No Calibration Error (noc)} & \multicolumn{3}{c}{Calibration Error (c)} \\
 Parameter & Fs-noc & FFD-noc & JSD & Fs-c & FFD-c & JSD \\
\hline
$t_c-t_0 [\mathrm{ms}]$ & $1.40^{+2.19}_{-3.31}$ & $1.58^{+1.99}_{-2.93}$ & 0.014 & $1.11^{+2.50}_{-3.85}$ & $1.36^{+2.27}_{-3.67}$ & 0.017 \\
$\mathcal{M} [M_{\odot}]$ & $6.51^{+0.44}_{-0.45}$ & $6.52^{+0.41}_{-0.44}$ & 0.004 & $6.51^{+0.44}_{-0.46}$ & $6.52^{+0.41}_{-0.45}$ & 0.005 \\
$q$ & $0.19^{+0.63}_{-0.14}$ & $0.20^{+0.58}_{-0.14}$ & 0.004 & $0.19^{+0.64}_{-0.13}$ & $0.20^{+0.60}_{-0.14}$ & 0.004 \\
$\chi_1$ & $0.00^{+0.16}_{-0.13}$ & $-0.01^{+0.09}_{-0.11}$ & 0.052 & $0.01^{+0.22}_{-0.15}$ & $-0.01^{+0.13}_{-0.13}$ & 0.047 \\
$\chi_2$ & $0.00^{+0.42}_{-0.40}$ & $-0.01^{+0.40}_{-0.40}$ & 0.008 & $0.01^{+0.46}_{-0.44}$ & $0.01^{+0.46}_{-0.42}$ & 0.011 \\
$\theta_{JN} [\mathrm{rad}]$ & $0.97^{+1.57}_{-0.66}$ & $0.66^{+1.74}_{-0.45}$ & 0.105 & $0.99^{+1.59}_{-0.68}$ & $0.71^{+1.93}_{-0.50}$ & 0.075 \\
$\phi [\mathrm{rad}]$ & $3.13^{+2.78}_{-2.78}$ & $2.98^{+2.75}_{-2.49}$ & 0.007 & $3.16^{+2.80}_{-2.84}$ & $2.65^{+3.05}_{-2.28}$ & 0.014 \\
$\alpha [\mathrm{rad}]$ & $0.23^{+0.18}_{-0.04}$ & $0.22^{+0.16}_{-0.03}$ & 0.020 & $0.23^{+0.18}_{-0.04}$ & $0.23^{+0.18}_{-0.03}$ & 0.002 \\
$\delta [\mathrm{rad}]$ & $-0.44^{+0.04}_{-0.13}$ & $-0.43^{+0.04}_{-0.12}$ & 0.019 & $-0.44^{+0.04}_{-0.13}$ & $-0.44^{+0.04}_{-0.13}$ & 0.004 \\
$d_{L} [\mathrm{Mpc}]$ & $326.0^{+53.8}_{-104.0}$ & $303.9^{+68.5}_{-104.0}$ & 0.027 & $324.0^{+55.6}_{-101.9}$ & $304.4^{+68.4}_{-115.4}$ & 0.031 \\
$\psi [\mathrm{rad}]$ & $1.64^{+1.34}_{-1.48}$ & $1.60^{+1.26}_{-1.36}$ & 0.011 & $1.54^{+1.40}_{-1.38}$ & $1.48^{+1.37}_{-1.29}$ & 0.012 \\
$\lambda_{2}$ & $2559.6^{+2157.9}_{-2263.2}$ & $2463.1^{+2067.5}_{-2096.0}$ & 0.008 & $2526.9^{+2191.9}_{-2250.0}$ & $2564.7^{+2057.9}_{-2228.3}$ & 0.004 \\
\hline
$N_s$ & $4921$ & $5106$ & - & $5660$ & $5856$ & - \\
$T_s [h]$ & $9.7$ & $12.8$ & - & $17.9$ & $40.8$ & - \\
$\ln\mathcal{Z}$ & $237.5$ & $237.3$ & - & $237.2$ & $236.9$ & - \\
\end{tabular}
\end{ruledtabular}
\caption{
Parameters of GW190814 inferred using the \texttt{IMRPhenomNSBH} waveform model under the NSBH hypothesis, estimated with the $\fs$ (Fs) and FFD methods.
Results are shown for analyses without calibration uncertainty (noc) and with calibration uncertainty (c), including the tidal deformability parameter of the secondary object, $\lambda_2$.
The \ac{JSD} quantifies the agreement between the posterior distributions obtained with the two methods. The table also reports the number of posterior samples ($N_s$), sampling time ($T_s$), and log Bayesian evidence ($\ln\mathcal{Z}$) for each analysis.
}
\label{tab:GW190814nsbh_allp}
\begin{tablenotes}
    \small
    \item[a] 1. The geocentric time $t_c$ is reported as an offset from the GW190814 trigger time, $t_0=1249852256.99$ s.
\end{tablenotes}
\end{table*}

An important methodological observation is the role of calibration uncertainty in mitigating the effects of likelihood multimodality. As shown in Fig.~\ref{fig:GW190814bbh_lls}, including calibration uncertainty systematically reduces the relative prominence of the secondary likelihood peak in the $\fs$ analysis. This improved robustness is also reflected quantitatively in Table.~\ref{tab:GW190814xphmST_allp}. For the majority of parameters, the \ac{JSD} between the $\fs$ and FFD posteriors decreases when calibration uncertainty is included, indicating better consistency between the two inference methods. For example, the JSD for the primary spin magnitude $a_1$ drops from $0.034$ to $0.012$, while for the spin phase angle $\phi_{12}$ it decreases from $0.045$ to $0.016$. Similarly, the agreement for luminosity distance $d_L$ and polarization angle $\psi$ improves, with JSDs falling from $0.009$ to $0.005$ and $0.011$ to $0.004$, respectively.

In terms of computational performance, the $\fs$ method retains its efficiency advantage. In the absence of calibration errors, the $\fs$ analysis completed in $34.3$\,hours, compared to $44.1$\,hours for the FFD method---a time saving of about $22\%$. Although this speed-up is less pronounced than in other analyses in this work, likely due to the additional sampling required to explore the bimodal likelihood landscape, the advantage of the $\fs$ method becomes more significant when calibration uncertainty is included. In this more complex scenario, the $\fs$ method took $59.6$\,hours versus $95.6$\,hours for the FFD, increasing the time saving to approximately $38\%$. Furthermore, the $\fs$ analysis yields a higher Bayesian evidence in both scenarios, suggesting a better exploration of the parameter space. This demonstrates that even for challenging signals with complex posterior structures, the $\fs$ method provides a notable and increasingly beneficial reduction in computational cost.

\subsubsection{Results of IMRPhenomNSBH}
While the preceding analysis treated GW190814 as a BBH coalescence, the true nature of the secondary component remains ambiguous. With a mass of approximately $2.6\Msun$, this object sits within the lower mass gap, potentially representing either the lightest black hole or the heaviest neutron star ever observed in a compact binary system. Given this uncertainty, it is essential to investigate the signal under the hypothesis that the secondary is a neutron star, which introduces the possibility of tidal interactions. To this end, we repeat our analysis using the \texttt{IMRPhenomNSBH} waveform approximant, which includes the tidal deformability parameter ($\lambda_2$) for the secondary object. In this section, we present the results of this NSBH-specific analysis. We begin by illustrating the joint posterior distributions of the luminosity distance and polarization angle to facilitate comparison with the BBH models discussed previously. Following this, we report the detailed constraints on the full set of source parameters inferred from the \texttt{IMRPhenomNSBH} template.

\begin{figure*}
\centering
\begin{subfigure}[b]{0.88\linewidth}
\centering
\includegraphics[width=\textwidth,height=12cm]{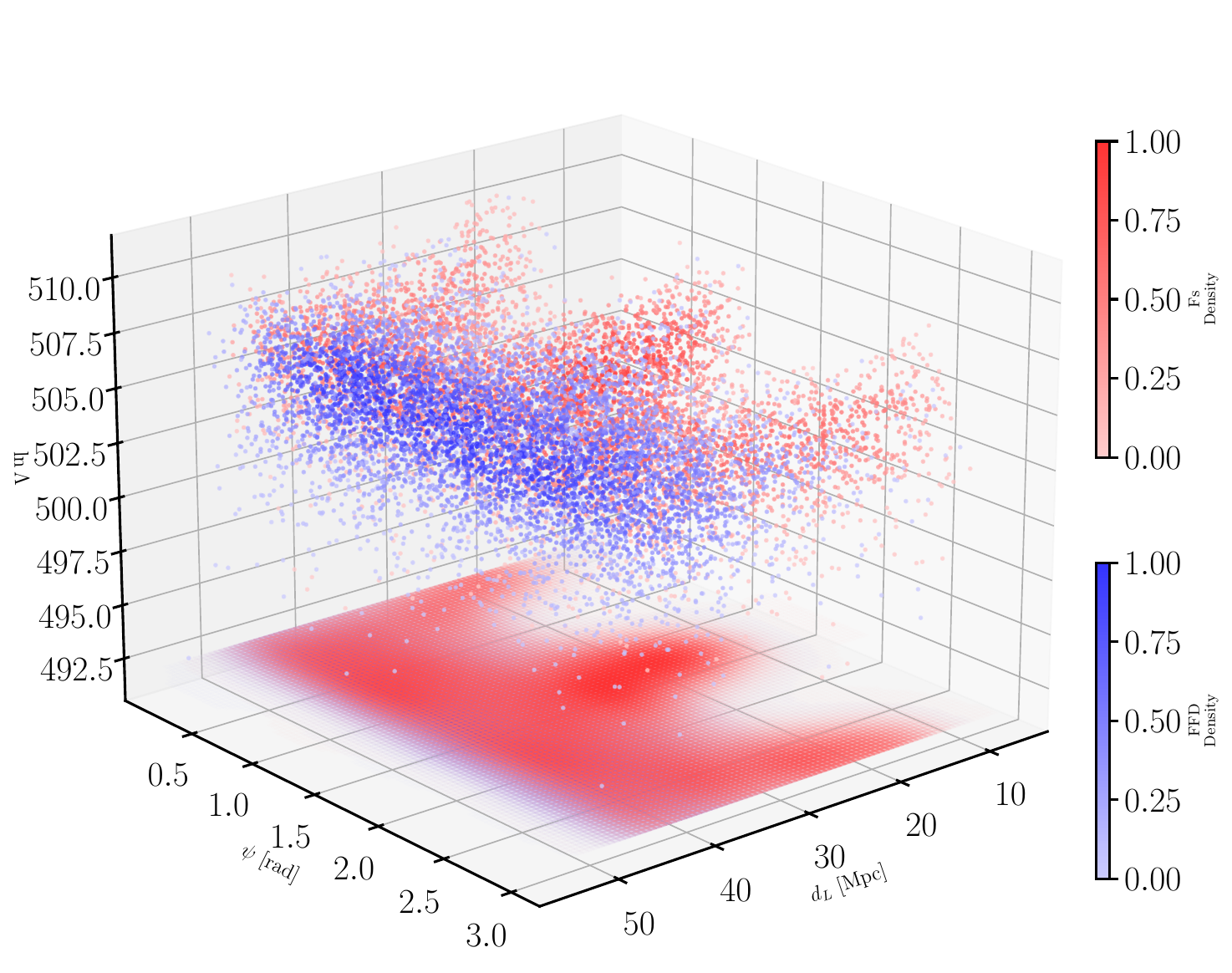}
\end{subfigure}\\
\begin{subfigure}[b]{0.48\linewidth}
\centering
\includegraphics[width=\textwidth,height=9cm]{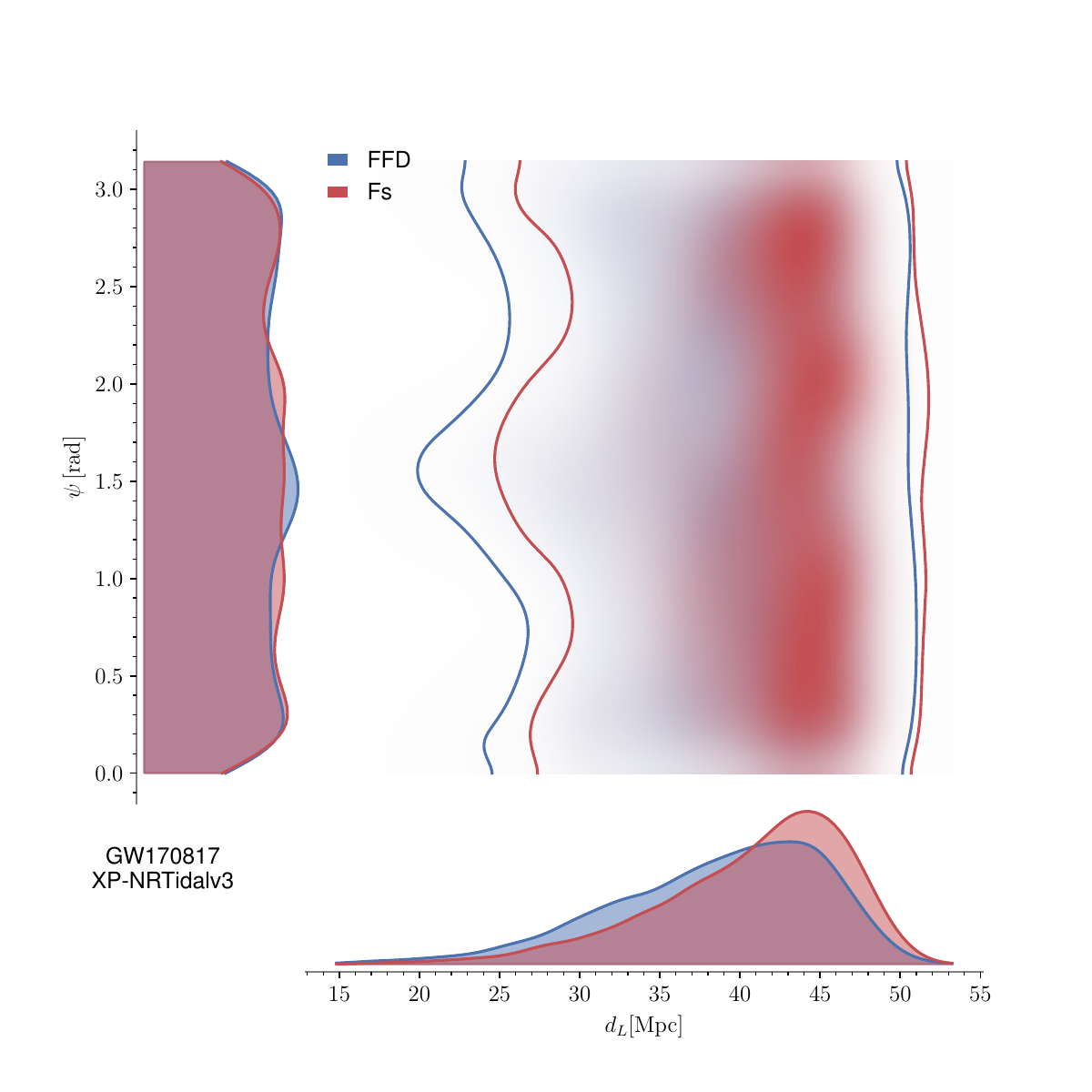}
\end{subfigure}%
\begin{subfigure}[b]{0.48\linewidth}
\centering
\includegraphics[width=\textwidth,height=9cm]{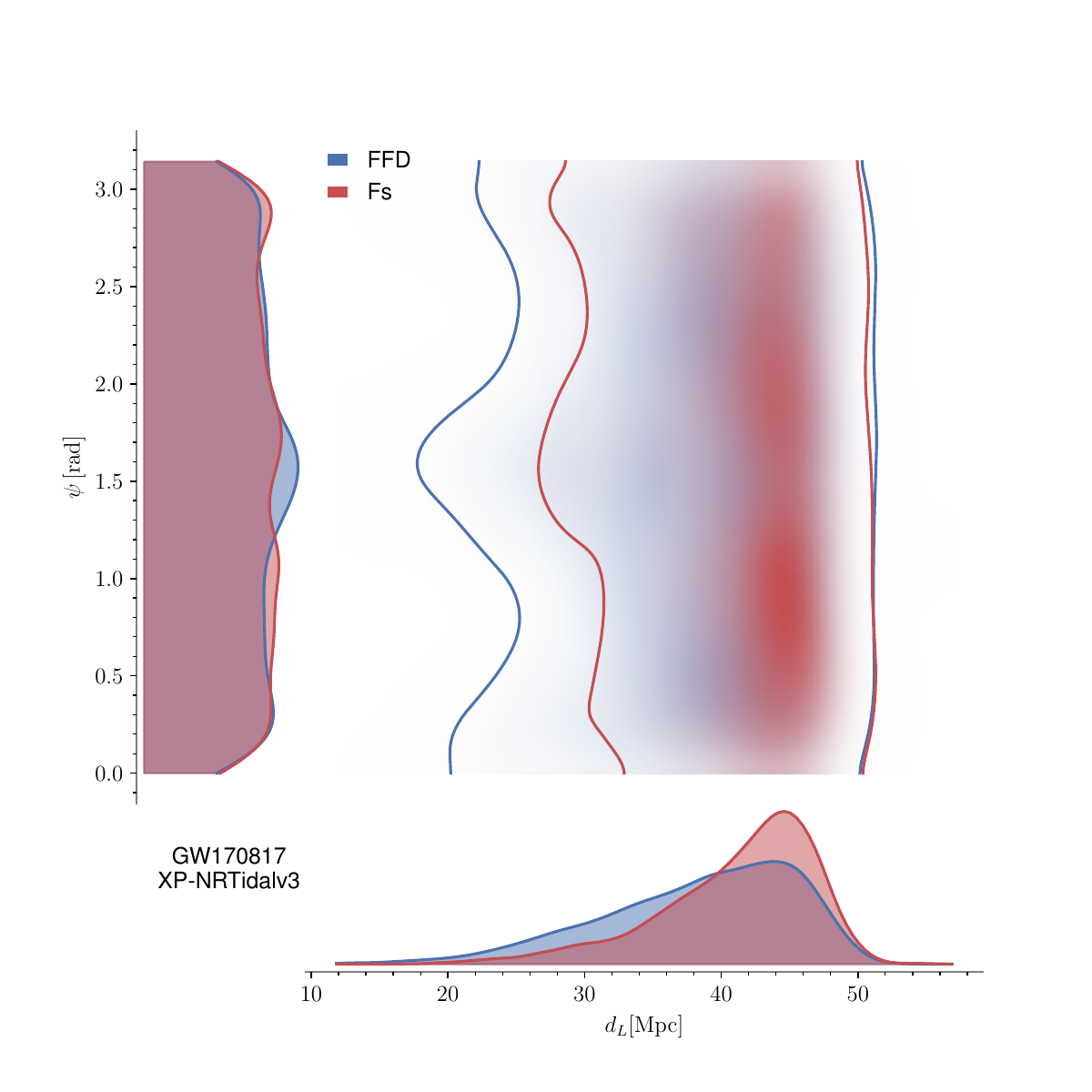}
\end{subfigure}%
\caption{
Same as Fig.~\ref{fig:dl-psi1}, but for the analysis of GW170817 under the binary neutron star hypothesis using the \texttt{IMRPhenomXP-NRTidalv3} waveform model.
}
\label{fig:dl-psi4}
\end{figure*}

Fig.~\ref{fig:dl-psi3} presents the joint posterior distributions of the luminosity distance and polarization angle for GW190814 under the NSBH hypothesis, inferred using the \texttt{IMRPhenomNSBH} waveform model.
For both analyses with and without calibration uncertainty, the $\fs$ and FFD methods yield highly consistent posterior structures, with nearly identical support regions in the $(d_L, \psi)$ plane. No additional multimodality is observed compared to the BBH-based analyses discussed previously.

\begin{table*}[hptb]
\begin{ruledtabular}
\begin{tabular}{c | c c c | c c c}
 & \multicolumn{3}{c|}{No Calibration Error (noc)} & \multicolumn{3}{c}{Calibration Error (c)} \\
 Parameter & Fs-noc & FFD-noc & JSD & Fs-c & FFD-c & JSD \\
\hline
$t_c-t_0 [\mathrm{ms}]$ & $-1.97^{+1.60}_{-3.15}$ & $-2.27^{+1.62}_{-3.25}$ & 0.015 & $-2.12^{+1.88}_{-3.21}$ & $-2.33^{+1.75}_{-3.34}$ & 0.012 \\
$\mathcal{M}-\mathcal{M}_0 [10^{-4}M_{\odot}]$ & $5.9^{+1.5}_{-1.3}$ & $6.0^{+1.5}_{-1.3}$ & 0.002 & $6.1^{+1.6}_{-1.4}$ & $6.0^{+1.5}_{-1.4}$ & 0.003 \\
$q$ & $0.87^{+0.12}_{-0.16}$ & $0.87^{+0.11}_{-0.14}$ & 0.008 & $0.87^{+0.11}_{-0.16}$ & $0.87^{+0.11}_{-0.14}$ & 0.004 \\
$a_{1}$ & $0.02^{+0.02}_{-0.02}$ & $0.02^{+0.02}_{-0.02}$ & 0.006 & $0.02^{+0.02}_{-0.02}$ & $0.02^{+0.02}_{-0.02}$ & 0.004 \\
$a_{2}$ & $0.02^{+0.02}_{-0.02}$ & $0.02^{+0.02}_{-0.02}$ & 0.008 & $0.02^{+0.02}_{-0.02}$ & $0.02^{+0.02}_{-0.02}$ & 0.004 \\
$\theta_{1} [\mathrm{rad}]$ & $1.26^{+1.10}_{-0.88}$ & $1.30^{+0.99}_{-0.80}$ & 0.010 & $1.26^{+1.13}_{-0.89}$ & $1.29^{+1.07}_{-0.87}$ & 0.004 \\
$\theta_{2} [\mathrm{rad}]$ & $1.37^{+1.12}_{-0.95}$ & $1.33^{+1.01}_{-0.85}$ & 0.007 & $1.32^{+1.13}_{-0.93}$ & $1.33^{+1.06}_{-0.90}$ & 0.003 \\
$\phi_{12} [\mathrm{rad}]$ & $3.08^{+2.88}_{-2.77}$ & $3.14^{+2.83}_{-2.84}$ & 0.002 & $3.15^{+2.86}_{-2.85}$ & $3.16^{+2.81}_{-2.87}$ & 0.003 \\
$\phi_{JL} [\mathrm{rad}]$ & $3.18^{+2.80}_{-2.88}$ & $3.22^{+2.76}_{-2.90}$ & 0.003 & $3.16^{+2.83}_{-2.83}$ & $3.06^{+2.93}_{-2.76}$ & 0.003 \\
$\theta_{JN} [\mathrm{rad}]$ & $2.30^{+0.58}_{-0.37}$ & $2.54^{+0.41}_{-0.43}$ & 0.103 & $2.34^{+0.56}_{-0.41}$ & $2.54^{+0.42}_{-0.48}$ & 0.060 \\
$\phi [\mathrm{rad}]$ & $3.13^{+2.72}_{-2.82}$ & $3.15^{+2.78}_{-2.83}$ & 0.009 & $3.18^{+2.71}_{-2.85}$ & $3.19^{+2.73}_{-2.88}$ & 0.008 \\
$\alpha [\mathrm{rad}]$ & $3.42^{+0.04}_{-0.04}$ & $3.43^{+0.04}_{-0.04}$ & 0.008 & $3.43^{+0.04}_{-0.04}$ & $3.43^{+0.04}_{-0.04}$ & 0.007 \\
$\delta [\mathrm{rad}]$ & $-0.37^{+0.06}_{-0.06}$ & $-0.37^{+0.06}_{-0.07}$ & 0.004 & $-0.37^{+0.07}_{-0.06}$ & $-0.37^{+0.07}_{-0.06}$ & 0.004 \\
$d_{L} [\mathrm{Mpc}]$ & $42.0^{+6.3}_{-12.8}$ & $39.6^{+7.8}_{-13.1}$ & 0.025 & $42.5^{+5.6}_{-12.8}$ & $39.5^{+8.1}_{-14.4}$ & 0.038 \\
$\psi [\mathrm{rad}]$ & $1.54^{+1.47}_{-1.40}$ & $1.54^{+1.48}_{-1.42}$ & 0.004 & $1.55^{+1.45}_{-1.44}$ & $1.56^{+1.45}_{-1.42}$ & 0.005 \\
$\lambda_{1}$ & $319.0^{+831.0}_{-290.4}$ & $375.3^{+757.0}_{-326.6}$ & 0.011 & $351.5^{+880.8}_{-319.8}$ & $388.4^{+783.9}_{-341.8}$ & 0.005 \\
$\lambda_{2}$ & $525.9^{+1108.3}_{-474.6}$ & $544.8^{+919.5}_{-470.4}$ & 0.008 & $589.5^{+1227.6}_{-525.7}$ & $566.5^{+942.2}_{-495.1}$ & 0.010 \\
\hline
$N_s$ & $5267$ & $5785$ & - & $5344$ & $5863$ & - \\
$T_s [h]$ & $17.0$ & $35.6$ & - & $16.6$ & $84.2$ & - \\
$\ln\mathcal{Z}$ & $474.7$ & $474.1$ & - & $473.6$ & $473.3$ & - \\
\end{tabular}
\end{ruledtabular}
\caption{Parameters of GW170817 estimated using \texttt{IMRPhenomXP-NRTidalv3} waveform model by $\fs$ (Fs) and FFD methods. The table compares results without calibration error (noc) and with calibration error (c), including tidal deformability parameters $\lambda_1$ and $\lambda_2$. JSD values, sample size ($N_s$), sampling time ($T_s$), and log Bayes factor ($\ln\mathcal{Z}$) are included.}
\label{tab:GW170817_allp}
\begin{tablenotes}
    \small
    \item[a] 1. The geocentric time $t_c$ is reported as an offset from the GW170817 trigger time, $t_0=1187008882.43$ s.
	\item[b] 2. The redshifted chirp mass $\mathcal{M}$ is reported as an offset from the reference mass, $\mathcal{M}_0=1.197\Msun$.
\end{tablenotes}
\end{table*}

The quantitative agreement between the two inference approaches is further confirmed in Table~\ref{tab:GW190814nsbh_allp}. For most intrinsic and extrinsic parameters, including the tidal deformability $\lambda_2$, the \ac{JSD} values remain small, indicating a high degree of consistency. However, we note a notable exception for the inclination angle $\theta_{JN}$, where the JSD is $0.105$ without calibration uncertainty. This discrepancy likely reflects the limitations of the \texttt{IMRPhenomNSBH} model, which does not account for the spin-precession and higher-order mode effects that are significant for GW190814, in contrast to the more consistent results obtained with the \texttt{IMRPhenomXPHM-SpinTaylor} waveform. This highlights the critical importance of using waveform models that fully capture the signal's physical content. Nevertheless, including calibration uncertainty mitigates this disagreement, reducing the JSD for $\theta_{JN}$ to $0.075$, underscoring the importance of accounting for instrumental effects.

In addition to statistical consistency, the $\fs$ method demonstrates a significant advantage in computational efficiency. As shown in Table~\ref{tab:GW190814nsbh_allp}, without calibration uncertainty, the $\fs$ analysis required $9.7$ hours compared to $12.8$ hours for the FFD approach, a time saving of approximately $24\%$. This efficiency gain becomes substantially more pronounced when calibration is included: the $\fs$ analysis completed in $17.9$ hours, while the FFD method required $40.8$ hours, representing a speed-up factor of more than two (a time saving of $\sim 56\%$). As with the other analyses, the $\fs$ method yields a slightly higher Bayesian evidence, further supporting its robustness. This highlights the practical advantage of the $\fs$ method for computationally intensive IMR analyses involving complex waveform models and extended parameter spaces.

Overall, these results demonstrate that the $\fs$ method provides a computationally efficient yet statistically consistent alternative to FFD inference for NSBH analyses of GW190814, even in the presence of calibration uncertainty and tidal parameters.

\subsection{Analysis of GW170817: A Binary Neutron Star Merger}
Following the BBH and NSBH analyses presented above, we next consider a binary neutron star system to examine the broader applicability of the $\mathcal{F}$-statistic–based inference framework.
We choose GW170817, the first observed BNS merger, as a representative and well-controlled test case. Owing to its high signal-to-noise ratio, long inspiral duration in the sensitive band, and well-established astrophysical interpretation, GW170817 provides an ideal benchmark for validating inference strategies beyond black-hole–dominated systems.

This analysis is designed to demonstrate that the $\mathcal{F}$-statistic approach is applicable across the full range of compact binary coalescences considered in this work, complementing the BBH and NSBH studies presented above. We perform parameter estimation using the \texttt{IMRPhenomXP-NRTidalv3} waveform model and adopt the same low-spin prior as used in the LVK analyses of GW170817. As in the previous sections, we compare the $\mathcal{F}$-statistic–based method with the standard Bayesian inference both with and without calibration uncertainty, allowing for a consistent assessment of the impact of calibration errors in the BNS regime.

Fig.~\ref{fig:dl-psi4} and Table~\ref{tab:GW170817_allp} summarize the parameter-estimation results for GW170817 obtained with the $\mathcal{F}$-statistic–based and FFD likelihood methods, both without and with calibration uncertainty. Across the full set of intrinsic and extrinsic parameters, the two inference approaches yield statistically consistent posterior distributions. This consistency is quantitatively supported by the \ac{JSD} values in Table~\ref{tab:GW170817_allp}. For instance, while there is a modest tension in the inclination angle $\theta_{JN}$ without calibration ($\text{JSD}=0.103$), including calibration uncertainty improves the agreement significantly, reducing the JSD to $0.060$. Similar improvements are seen for other key parameters, such as the mass ratio $q$ (JSD drops from $0.008$ to $0.004$) and the primary tidal deformability $\lambda_1$ (JSD drops from $0.011$ to $0.005$), highlighting the importance of marginalizing over calibration errors for robust inference.

A significant practical difference between the two approaches emerges in their computational efficiency. As shown by the sampling times in Table~\ref{tab:GW170817_allp}, without calibration error, the $\fs$-based analysis required $17.0$ hours, while the FFD method took $35.6$ hours, corresponding to a speed-up factor of over two. This computational advantage becomes dramatically more pronounced when calibration uncertainty is included. In this more complex scenario, the FFD analysis time more than doubled to $84.2$ hours, whereas the $\fs$ analysis completed in just $16.6$ hours. This represents a speed-up factor of more than five, demonstrating that the efficiency gains of the $\fs$ method are even more critical for realistic analyses of long-duration signals that include instrumental uncertainties and extended physical models. As seen in Table~\ref{tab:GW170817_allp}, the $\fs$-based analysis consistently yields a higher Bayesian evidence than the FFD method.

Taken together, the GW170817 results demonstrate that the $\mathcal{F}$-statistic–based framework provides parameter estimates consistent with standard Bayesian inference, while offering a substantial reduction in computational cost. This confirms the robustness and efficiency of the method for binary neutron star systems, complementing the BBH and NSBH analyses presented above and establishing its applicability across a broad range of compact binary coalescences.

\subsection{Comparison with Luminosity-Distance Marginalization in {\sc Bilby}}\label{ssec:bilby_dl_marg}
In addition to the comparison between the \ac{FFD} and $\fs$ methods presented above, it is also useful to benchmark the latter against an existing acceleration strategy that is already widely used in practical \ac{GW} inference workflows, namely luminosity-distance marginalization in {\sc Bilby}. In the standard implementation, this procedure marginalizes numerically over the luminosity distance through a Riemann sum evaluated on a precomputed grid in matched-filter \ac{SNR} and optimal \ac{SNR}, with a lookup table used to accelerate interpolation \citep{Romero-Shaw:2020owr}. Since this treatment does not modify the likelihood function itself or the functional dependence among the remaining parameters, it is generally expected to preserve the parameter-estimation results of the full Bayesian analysis while often reducing the computational cost. At the same time, previous work has shown that the practical gain is not universal and may depend on the sampler configuration \citep{Williams:2021qyt}.

We further performed an additional set of analyses using {\sc Bilby} with luminosity-distance marginalization enabled. To ensure a controlled comparison, we considered the same events, waveform models, and calibration-uncertainty treatment adopted in the corresponding analyses discussed above. Because the consistency of posterior estimation under luminosity-distance marginalization has already been examined in earlier studies and is not the main focus of the present paper, we restrict the discussion here to practical efficiency and evidence behavior.

\begin{table*}[htbp]
\begin{ruledtabular}
\begin{tabular}{lccc}
Event and source/model type & $N_s$ & $T_s\,[\mathrm{h}]$ & $\ln \mathcal{Z}$ \\
\hline
GW190412 (BBH, \texttt{IMRPhenomXPHM-SpinTaylor}) & 6993 & 17.9 & 146.2 \\
GW190814 (BBH, \texttt{IMRPhenomXPHM-SpinTaylor}) & 6840 & 99.1 & 258.7 \\
GW190814 (NSBH, \texttt{IMRPhenomNSBH}) & 5799 & 37.1 & 236.9 \\
GW170817 (BNS, \texttt{IMRPhenomXP-NRTidalv3}) & 5624 & 60.2 & 473.3 \\
\end{tabular}
\caption{Summary of the additional analyses performed with luminosity-distance marginalization enabled in {\sc Bilby}, including the number of posterior samples $N_s$, the total sampling time $T_s$, and the log Bayesian evidence $\ln\mathcal{Z}$.}
\label{tab:dlmarg_summary}
\end{ruledtabular}
\end{table*}

The results are summarized in Table~\ref{tab:dlmarg_summary}. When compared with the corresponding full Bayesian analyses reported earlier, the behavior of luminosity-distance marginalization is found to be source dependent. For GW190814 analyzed with \texttt{IMRPhenomNSBH} and for GW170817 analyzed with \texttt{IMRPhenomXP-NRTidalv3}, the outcome is most consistent with the usual expectation: the differences in log Bayesian evidence are negligible, and the total sampling time is reduced relative to the corresponding \ac{FFD} runs. These cases therefore confirm that luminosity-distance marginalization can provide a practical speed-up while preserving the overall statistical behavior of the inference.

By contrast, the cases based on \texttt{IMRPhenomXPHM-SpinTaylor} show a less uniform pattern. For both GW190412 and GW190814 under the \ac{BBH} hypothesis, the differences in log evidence with respect to the corresponding \ac{FFD} analyses are modest but visible, at the level of about $0.4$. A plausible explanation is that these signals involve more complicated likelihood envelopes because of the combined effects of spin precession and higher-order modes, which may limit the extent to which standard distance marginalization alone improves the sampling performance.

This trend is particularly evident for GW190814 analyzed with \texttt{IMRPhenomXPHM-SpinTaylor}. In that case, enabling luminosity-distance marginalization does not reduce the total sampling time. We also note that this is the only case in which the number of posterior samples increases relative to the corresponding \ac{FFD} analysis. Although this may suggest some connection between wall-clock time and sampling behavior in this particular case, we do not attempt to draw a strong conclusion, since the final computational cost can also depend on waveform-evaluation expense and on the detailed behavior of the sampler. A more systematic investigation of these effects would be interesting, but it lies beyond the scope of the present work.

For comparison, the $\fs$ analysis for the same GW190814+\texttt{IMRPhenomXPHM-SpinTaylor} case also yields a larger number of posterior samples than the corresponding \ac{FFD} analysis, yet it still remains substantially more efficient in total sampling time. This difference may reflect the fact that the $\fs$ method analytically maximizes over both the luminosity distance and the polarization angle in likelihood space, whereas the standard luminosity-distance-marginalization strategy removes only one of these two linear extrinsic parameters from the numerical sampling. In this sense, the present comparison further supports the practical advantage of the $\fs$ framework, especially for inference problems in which the likelihood structure is complicated and the cost of sampling extrinsic parameters remains non-negligible.

\section{Summary and Conclusion}\label{sec:con}
In this study, we conducted a comprehensive investigation of the $\fs$ method for parameter estimation of complete \ac{IMR} signals from a range of \ac{GW} sources. Our findings demonstrate that the $\fs$ offers significant advantages in performance, stability, and computational efficiency over the traditional \ac{FFD} approach. We presented a self-contained framework for the $\fs$ analysis that includes a method to reconstruct the posterior distributions for analytically maximized parameters under a physical prior, and a novel formulation to calculate the Bayesian evidence for robust model comparison.

Our comparative analysis across several landmark events—GW190412, GW190814, and GW170817—highlights several key findings. First, the $\fs$ method consistently produces posterior distributions that are in good agreement with those from standard FFD inference. Including calibration uncertainty tends to improve this agreement, suggesting that marginalization over instrumental effects helps harmonize the two approaches. Second, the $\fs$ method is computationally superior in the relative sense that, for each event and waveform configuration considered here, it requires less sampling time than the corresponding FFD analysis. This relative time saving is generally larger when calibration uncertainties are included, although the absolute wall-clock time of an individual run need not vary monotonically because it also depends on stochastic sampler behavior, convergence criteria, waveform-evaluation cost, and the detailed posterior geometry. Third, under the assumption of physical priors, the $\fs$-based analyses consistently yield higher Bayesian evidence than the corresponding FFD analyses. While this systematic difference is intrinsically linked to the analytical maximization over two parameters, which allows for a more efficient exploration of high-likelihood regions, we caution that such evidence comparisons are most safely interpreted within the same inference framework. While the $\fs$ method produced slightly broader posteriors for some parameters, we interpret this as a more honest and conservative quantification of statistical uncertainty, especially in complex, high-dimensional problems.

Overall, our study demonstrates the efficiency and reliability of the $\fs$ method. As detector sensitivities improve and GW detection rates increase, the need for efficient and robust analysis methods will become paramount. The advantages demonstrated here establish the $\fs$ as a promising tool for the rapid and reliable analysis of large event catalogs \citep{LIGOScientific:2025slb}. Future work will focus on applying this framework to a wider range of GW events, including those with lower \acp{SNR} and more complex physical features, and on testing its performance.

\begin{acknowledgments}
We thank the anonymous referee for their insightful comments.
H.-T. Wang is supported by ``the Natural Science Foundation of Liaoning Province" (Grant No. ZX20250217) and ``the Fundamental Research Funds for the Central Universities" at Dalian University of Technology.
This research has made use of data or software obtained from the Gravitational Wave Open Science Center~\cite{gwosc-url}, a service of LIGO Laboratory, the LIGO Scientific Collaboration, the Virgo Collaboration, and KAGRA~\cite{KAGRA:2023pio}. 
LIGO Laboratory and Advanced LIGO are funded by the United States National Science Foundation (NSF) as well as the Science and Technology Facilities Council (STFC) of the United Kingdom, the Max-Planck-Society (MPS), and the State of Niedersachsen/Germany for support of the construction of Advanced LIGO and construction and operation of the GEO600 detector. 
Additional support for Advanced LIGO was provided by the Australian Research Council. 
Virgo is funded, through the European Gravitational Observatory (EGO), by the French Centre National de Recherche Scientifique (CNRS), the Italian Istituto Nazionale di Fisica Nucleare (INFN) and the Dutch Nikhef, with contributions by institutions from Belgium, Germany, Greece, Hungary, Ireland, Japan, Monaco, Poland, Portugal, Spain.
KAGRA is supported by Ministry of Education, Culture, Sports, Science and Technology (MEXT), Japan Society for the Promotion of Science (JSPS) in Japan; National Research Foundation (NRF) and Ministry of Science and ICT (MSIT) in Korea; Academia Sinica (AS) and National Science and Technology Council (NSTC) in Taiwan of China.
\end{acknowledgments}

\section*{Data availability}
The posterior samples and analysis results supporting the findings of this study are publicly available \cite{DataFsIMR150914}.

\appendix
\section{Benchmark analysis: GW150914}\label{apsec:gw150914}
As a benchmark for our methodology, we include a re-analysis of the public data for the GW150914 event. This event serves as a well-understood, high-\ac{SNR} test case for validating inference methods. For this analysis, we used the \texttt{IMRPhenomXPHM} waveform model. The results presented here were obtained without including calibration uncertainty, allowing for a focused comparison of sampler performance and method stability, which complements the main text's discussion on calibration effects.

\subsection{Sampler Configuration}
The Bayesian inference was performed using the {\sc Bilby} package~\citep[v2.4.0;][]{Ashton:2018jfp}, with the nested sampling algorithm implemented in {\sc Dynesty}~\citep[v2.1.5;][]{Romero-Shaw:2020owr}.
To assess the stability and robustness of both the \ac{FFD} and $\fs$ methods, we employed two distinct sampler configurations, hereafter denoted `s1' and `s2'. Several key parameters were common to both configurations: the number of live points was set to $1000$, and the primary sampling method was ``rwalk''. The configurations differed in the following parameters:
\begin{itemize}
    \item \textbf{Configuration `s1':} The number of autocorrelation lengths was set to $50$, and the method for selecting new points was ``live''. This configuration is similar to the default settings used in the GWTC-2.1 and GWTC-3 catalog analyses.
    \item \textbf{Configuration `s2':} The number of autocorrelation lengths was reduced to $20$, proposals were drawn using a combination of ``diff'' and ``volumetric'' methods, and new points were selected using ``live-multi''. This alternative configuration was chosen for comparison, following standard practices outlined in the {\sc Bilby} documentation.
\end{itemize}
All analyses were parallelized using $50$ threads to reduce computational time.

\subsection{Performance and Stability Analysis}\label{apss:r3}
In this section, we present a detailed comparison of the parameter estimation results obtained from the $\fs$ and the \ac{FFD} methods for the GW150914 event. We evaluate the overall performance, efficiency, and stability of the two methods. We first compare their calculated Bayes factors and log-likelihood ratio distributions. Then, we examine posterior distributions for key parameter subsets: strongly degenerate extrinsic parameters ($t_c$, $\alpha$), the analytically maximized parameters ($d_L$, $\psi$), and finally all remaining source parameters. We assess stability across different sampler configurations using the \ac{JSD}.

\begin{figure}
\centering
\includegraphics[width=0.48\textwidth,height=6.6cm]{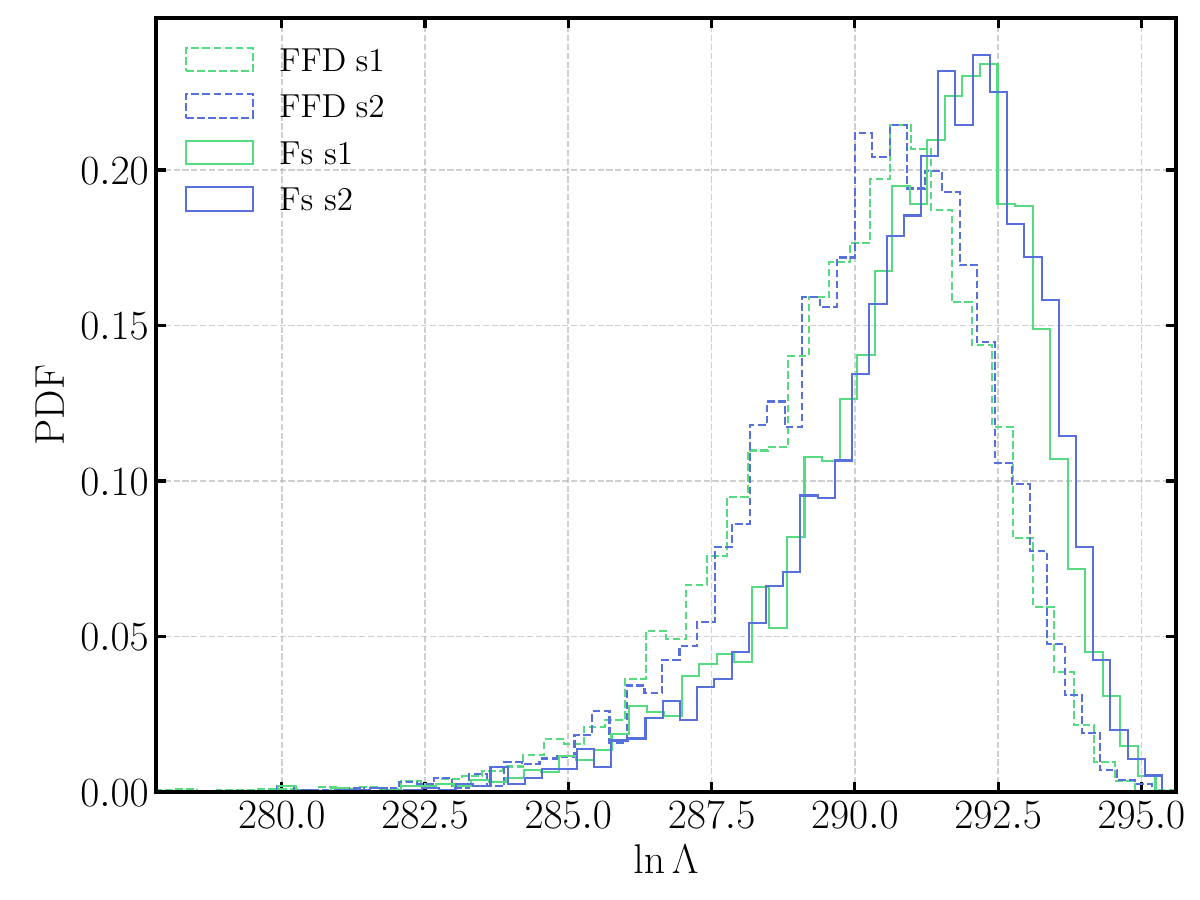}
\caption{
Posterior distributions of the log-likelihood ratio, $\ln\Lambda$, comparing the performance of the \ac{FFD} method (dashed histograms) and the $\fs$ method (solid histograms, denoted as `Fs'). The two colors represent different sampler configurations: `s1' (green) and `s2' (blue).
}\label{fig:ll12}
\end{figure}

\subsection{Model Evidence and Performance Comparison}
We begin the performance comparison by examining the posterior distributions of the log-likelihood ratio, $\ln\Lambda$, shown in Fig.~\ref{fig:ll12}. For both the \ac{FFD} and $\fs$ methods, the distributions are consistent across the two sampler settings (`s1' and `s2'). However, the distributions for the $\fs$ method are consistently shifted toward higher values, with median $\ln\Lambda$ values of $291.4$ (`s1') and $291.5$ (`s2'), compared to $290.3$ (`s1') and $290.4$ (`s2') for the \ac{FFD} runs. This suggests a better fit to the data, enabled by maximizing two linear parameters. 

For a more definitive and quantitative comparison, we compute the evidence, $\mathcal{Z}$, using the procedure outlined in Sec.~\ref{sec:methods}.
Then, the log-Bayes factor between two configurations of each method is defined as $\ln\mathcal{B}^{\rm s1}_{\rm s2} \equiv \ln\mathcal{Z}_{\rm s1} - \ln\mathcal{Z}_{\rm s2}$. 
The difference in log-evidence between the two sampler settings for the $\fs$ is only $0.1$, whereas for the \ac{FFD} method, the difference is $0.4$. This suggests that the $\fs$ method is more robust to changes in the sampler configuration for this event.

\begin{figure}
\centering
\includegraphics[width=0.5\textwidth,height=8.5cm]{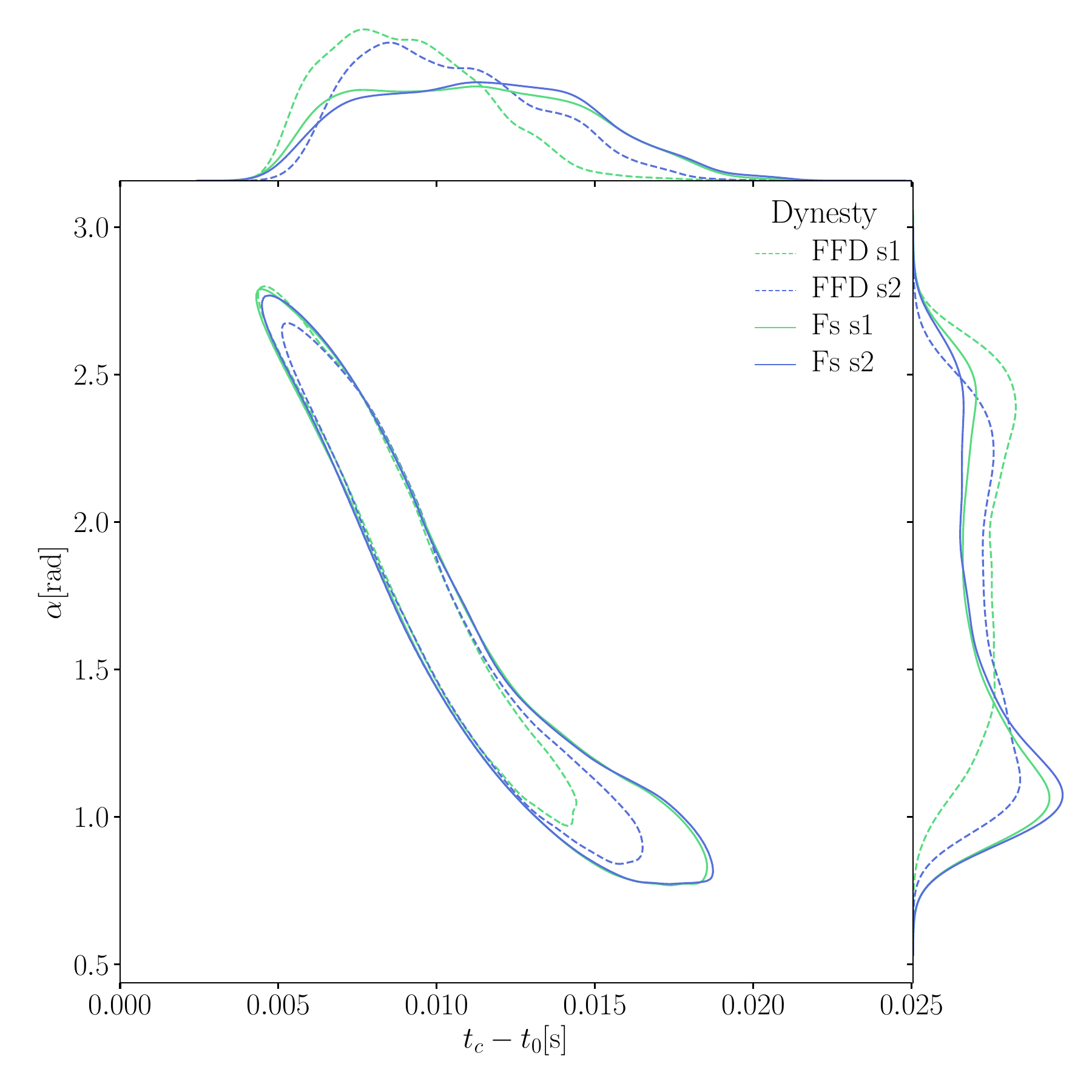}
\caption{
Comparison of the joint posterior distribution for the geocenter time, $t_c$, and right ascension, $\alpha$. The results from the \ac{FFD} method (dashed lines) and the $\fs$ method (solid lines, denoted as `Fs') are shown for two sampler configurations, `s1' (green) and `s2' (blue). The two-dimensional contours enclose the $90\%$ credible regions. For clarity, the geocenter time is shown as an offset from the event trigger time, $t_0=1126259462.4$ s. The top and right panels display the corresponding one-dimensional marginalized posterior distributions for each parameter.
}\label{fig:ra_tc1}
\end{figure}

\subsection{Constraints on Degenerate Extrinsic Parameters}
The geocenter time ($t_c$) and right ascension ($\alpha$) are crucial for localizing a gravitational wave source, but they often exhibit strong degeneracies that can challenge sampling algorithms. Fig.~\ref{fig:ra_tc1} compares the posterior distributions for these parameters obtained from both the \ac{FFD} and $\fs$ methods.

\begin{figure*}
\centering
\begin{subfigure}[b]{0.48\linewidth}
\centering
\includegraphics[width=\textwidth,height=6.6cm]{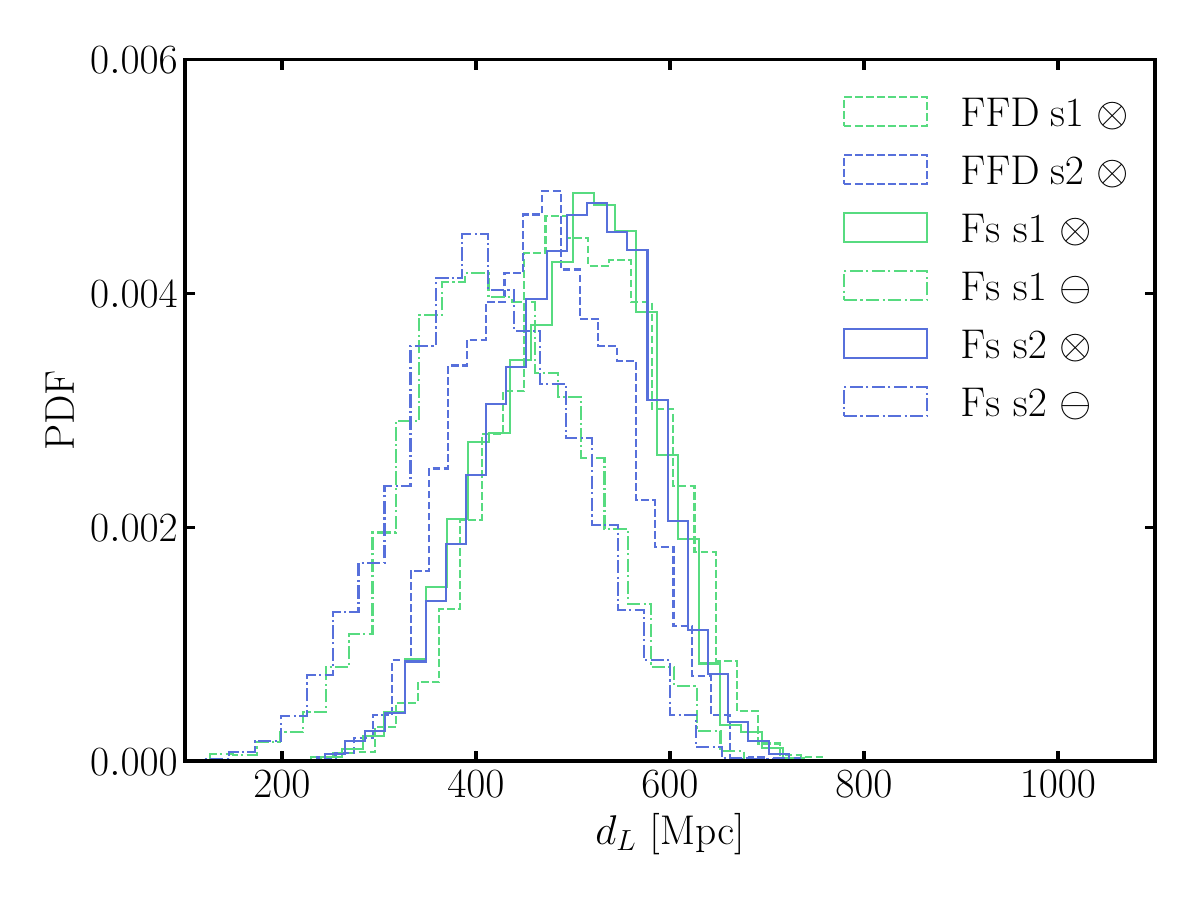}
\end{subfigure}%
\begin{subfigure}[b]{0.48\linewidth}
\centering
\includegraphics[width=\textwidth,height=6.6cm]{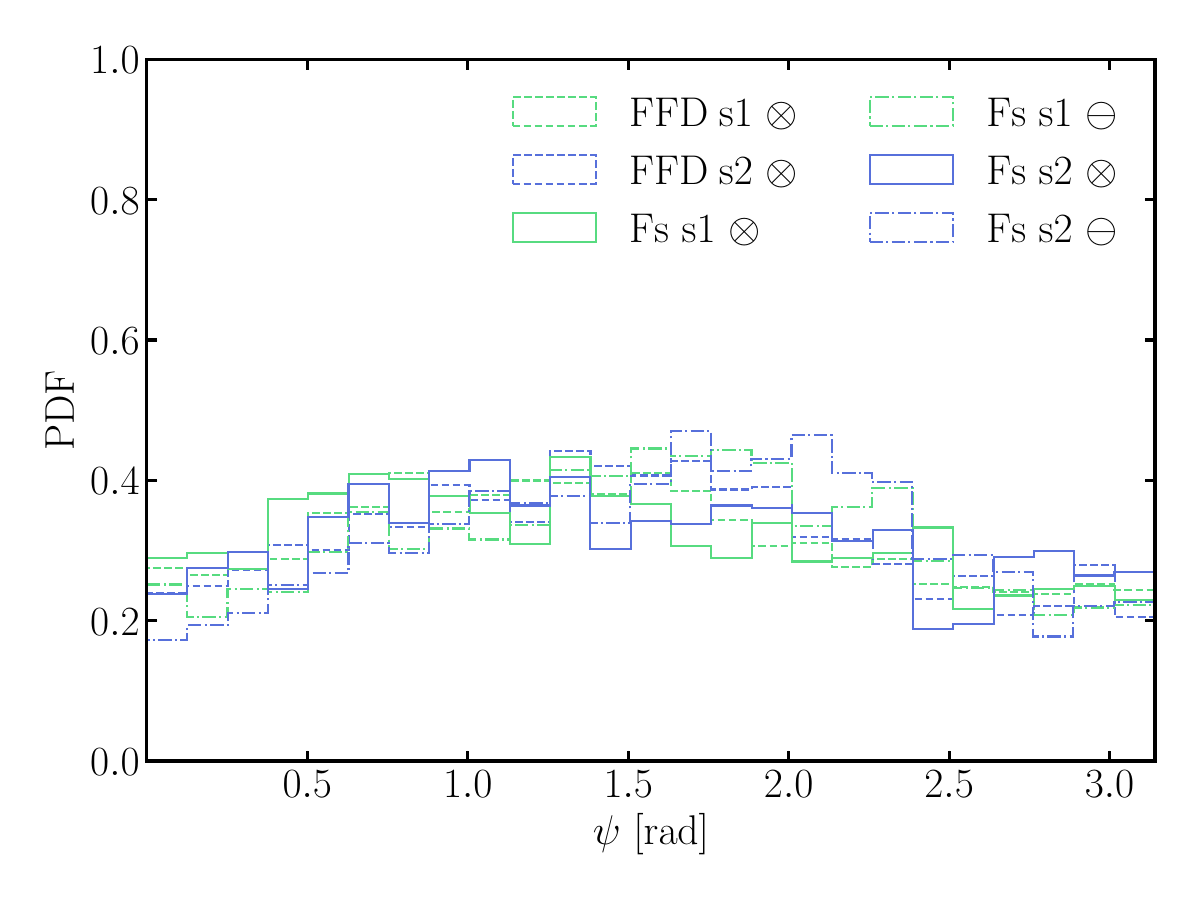}
\end{subfigure}%
\caption{
Comparison of the marginalized posterior distributions for the luminosity distance, $d_L$ (left panel), and the polarization angle, $\psi$ (right panel). The posteriors from the \ac{FFD} method are compared against the reconstructed posteriors from the $\fs$ (Fs) method for two sampler configurations (`s1' in green, `s2' in blue). For the $\fs$ results, posteriors are shown for two different prior assumptions: the initial reconstruction assuming a uniform prior on the linear parameters (marked $\ominus$), and the final result after re-weighting to the physical target prior (marked $\otimes$).
}\label{fig:dl-psi0}
\end{figure*}

We quantify the consistency between the `s1' and `s2' settings using the \ac{JSD}. 
A key finding is the superior stability of the $\fs$ method against changes in sampler configuration. We quantify this using the \ac{JSD} between the posteriors from the `s1' and `s2' settings. For the $\fs$ method, the JSD values are remarkably low: $0.004$ for $t_c$ and $0.005$ for $\alpha$. In contrast, the \ac{FFD} method yields significantly higher JSD values of $0.067$ for $t_c$ and $0.058$ for $\alpha$. This represents an order-of-magnitude improvement in consistency for the $\fs$ method.

The median values and $90\%$ credible intervals for each analysis further illustrate this point (where $t_0=1126259462.4$ s is the event trigger time):
\begin{itemize}
    \item \textbf{$\fs$ (`s1'):} $t_c = t_0+0.011^{+0.006}_{-0.005}$ s; $\alpha = 1.46^{+1.09}_{-0.52}$ rad.
    \item \textbf{$\fs$ (`s2'):} $t_c = t_0+0.011^{+0.006}_{-0.005}$ s; $\alpha = 1.40^{+1.11}_{-0.46}$ rad.
    \item \textbf{FFD (`s1'):} $t_c = t_0+0.009^{+0.004}_{-0.003}$ s; $\alpha = 1.91^{+0.66}_{-0.77}$ rad.
    \item \textbf{FFD (`s2'):} $t_c = t_0+0.010^{+0.005}_{-0.004}$ s; $\alpha = 1.63^{+0.81}_{-0.63}$ rad.
\end{itemize}
The consistency of the credible intervals from the $\fs$ method, supported by the low JSD values, demonstrates its enhanced robustness when characterizing these degenerate parameters.

\subsection{Reconstruction of Analytically Maximized Parameters}
The luminosity distance, $d_L$, and polarization angle, $\psi$, are treated fundamentally differently in the two frameworks. In the \ac{FFD} approach, they are sampled directly along with all other parameters. In the $\fs$ method, they are analytically maximized, and their posteriors must be reconstructed in a post-processing step, as detailed in Sec.~\ref{ssec:plp}. Fig.~\ref{fig:dl-psi0} shows the resulting posterior distributions from both methods using the physical target prior.

Despite the additional reconstruction step, the $\fs$ method demonstrates superior stability for the luminosity distance. 
For $d_L$, the JSD is only $0.003$ for the $\fs$ method, an order of magnitude smaller than the JSD of $0.043$ for the \ac{FFD} method. For the polarization angle $\psi$, the JSD values are low and comparable ($0.009$ for $\fs$ vs. $0.003$ for \ac{FFD}). This is expected, as $\psi$ is poorly constrained by the GW150914 event, leading to broad, featureless posteriors for both methods.

The median values and $90\%$ credible intervals from the physical target prior are:
\begin{itemize}
    \item \textbf{$\fs$ (`s1'):} $d_L = 504.8^{+115.0}_{-144.2}$ Mpc; $\psi = 1.41^{+1.52}_{-1.23}$ rad.
    \item \textbf{$\fs$ (`s2'):} $d_L = 502.0^{+117.0}_{-142.1}$ Mpc; $\psi = 1.48^{+1.47}_{-1.29}$ rad.
    \item \textbf{FFD (`s1'):} $d_L = 507.5^{+128.6}_{-134.1}$ Mpc; $\psi = 1.45^{+1.50}_{-1.27}$ rad.
    \item \textbf{FFD (`s2'):} $d_L = 469.7^{+130.3}_{-124.8}$ Mpc; $\psi = 1.50^{+1.43}_{-1.29}$ rad.
\end{itemize}
The greater consistency in the recovered constraints for $d_L$ indicates that the $\fs$ approach performs more reliably, even for parameters that are not sampled directly.

\subsection{Comparison of All Sampled Source Parameters}
\begin{figure*}[h]
\centering
\includegraphics[width=\textwidth,height=18cm]{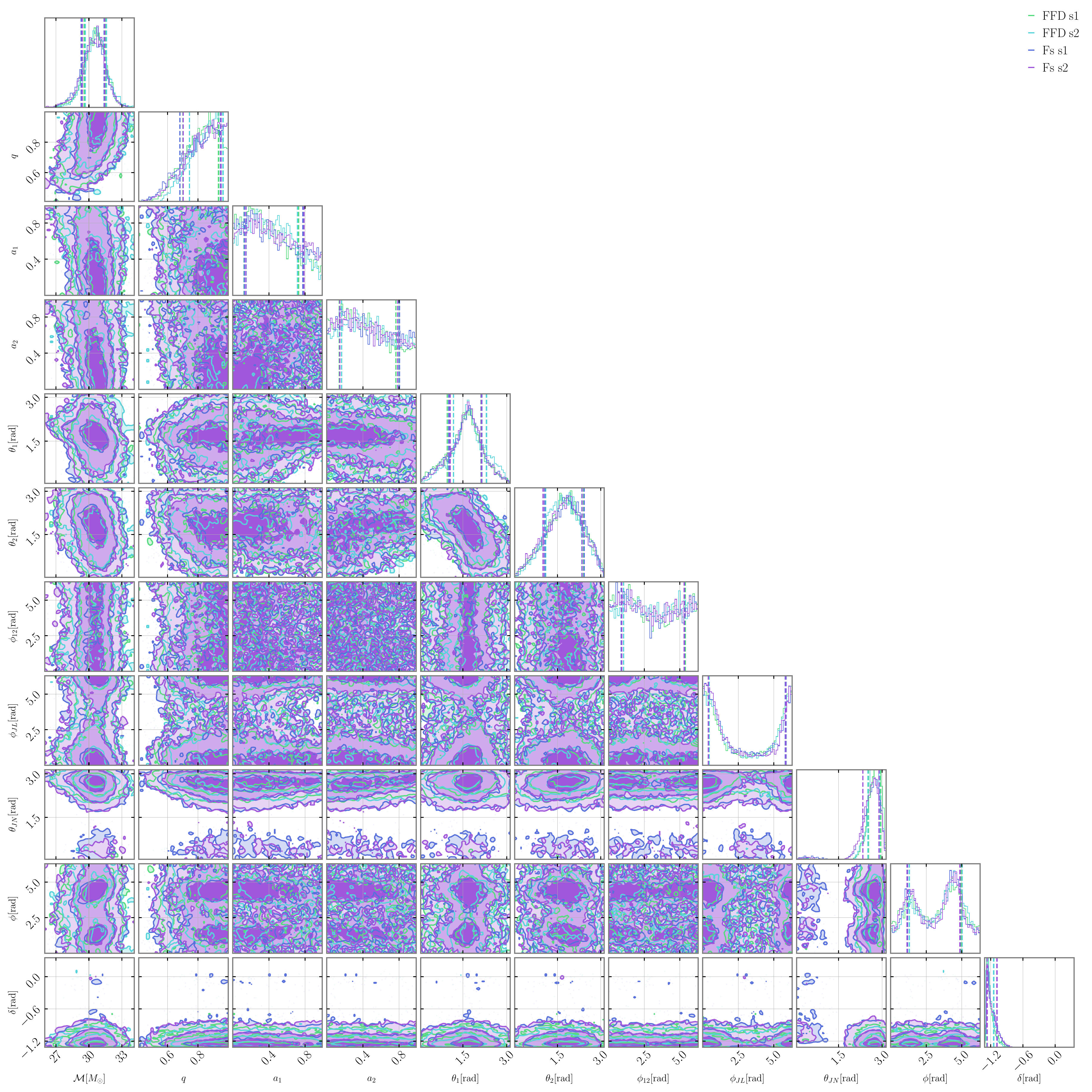}
\caption{
Full corner plot showing the posterior distributions for the intrinsic source parameters, comparing the results from the \ac{FFD} method and the $\fs$ (Fs) method under two different sampler settings (`s1' and `s2'). Two-dimensional contours indicate the $90\%$ credible regions, while the diagonal panels show the marginalized one-dimensional posterior distribution for each parameter.
}\label{fig:params2}
\end{figure*}

The posterior distributions for all remaining source parameters are shown in Fig.~\ref{fig:params2}, with the quantitative constraints summarized in Table~\ref{tab:params}. A primary finding from this comprehensive comparison is the enhanced stability of the $\fs$ method. As shown in Table~\ref{tab:params}, the \ac{JSD} values calculated between the `s1' and `s2' settings are systematically lower for the $\fs$ method across nearly all parameters, often by an order of magnitude. This demonstrates that the $\fs$ is significantly more robust to the specific configuration of the nested sampler.

\begin{table*}[hptb]
\begin{ruledtabular}
\begin{tabular}{l | c c c | c c c}
 & FFD s1 & FFD s2 & JSD & Fs s1 & Fs s2 & JSD \\
\hline
$t_{c}-t_0 [\mathrm{s}]$ & $0.009^{+0.004}_{-0.003}$ & $0.010^{+0.005}_{-0.004}$ & 0.067 & $0.011^{+0.006}_{-0.005}$ & $0.011^{+0.006}_{-0.005}$ & 0.004 \\
$\mathcal{M} [M_{\odot}]$ & $30.59^{+1.51}_{-1.80}$ & $30.66^{+1.63}_{-1.71}$ & 0.006 & $30.39^{+1.67}_{-1.99}$ & $30.50^{+1.56}_{-1.98}$ & 0.006 \\
$q$ & $0.83^{+0.15}_{-0.21}$ & $0.88^{+0.11}_{-0.22}$ & 0.026 & $0.84^{+0.15}_{-0.25}$ & $0.85^{+0.14}_{-0.24}$ & 0.005 \\
$a_{1}$ & $0.37^{+0.53}_{-0.33}$ & $0.42^{+0.46}_{-0.36}$ & 0.007 & $0.40^{+0.52}_{-0.36}$ & $0.43^{+0.49}_{-0.38}$ & 0.004 \\
$a_{2}$ & $0.41^{+0.51}_{-0.37}$ & $0.45^{+0.47}_{-0.40}$ & 0.005 & $0.43^{+0.50}_{-0.39}$ & $0.44^{+0.49}_{-0.39}$ & 0.004 \\
$\theta_{1} [\mathrm{rad}]$ & $1.61^{+0.91}_{-1.11}$ & $1.70^{+0.99}_{-0.99}$ & 0.017 & $1.65^{+0.88}_{-1.12}$ & $1.65^{+0.85}_{-1.02}$ & 0.003 \\
$\theta_{2} [\mathrm{rad}]$ & $1.76^{+0.98}_{-1.16}$ & $1.71^{+0.99}_{-1.01}$ & 0.007 & $1.79^{+0.96}_{-1.19}$ & $1.73^{+1.00}_{-1.18}$ & 0.004 \\
$\phi_{12} [\mathrm{rad}]$ & $2.92^{+3.06}_{-2.64}$ & $3.11^{+2.86}_{-2.80}$ & 0.004 & $2.89^{+3.10}_{-2.59}$ & $3.02^{+2.95}_{-2.71}$ & 0.004 \\
$\phi_{JL} [\mathrm{rad}]$ & $2.47^{+3.68}_{-2.35}$ & $2.23^{+3.94}_{-2.09}$ & 0.003 & $1.85^{+4.31}_{-1.71}$ & $1.67^{+4.48}_{-1.54}$ & 0.004 \\
$\theta_{JN} [\mathrm{rad}]$ & $2.77^{+0.27}_{-0.43}$ & $2.72^{+0.29}_{-0.40}$ & 0.013 & $2.66^{+0.35}_{-0.62}$ & $2.65^{+0.36}_{-0.57}$ & 0.006 \\
$\phi [\mathrm{rad}]$ & $3.40^{+2.38}_{-2.95}$ & $3.68^{+2.04}_{-3.14}$ & 0.005 & $3.47^{+2.22}_{-2.93}$ & $3.37^{+2.31}_{-2.85}$ & 0.003 \\
$\alpha [\mathrm{rad}]$ & $1.91^{+0.66}_{-0.77}$ & $1.63^{+0.81}_{-0.63}$ & 0.058 & $1.46^{+1.07}_{-0.52}$ & $1.40^{+1.11}_{-0.46}$ & 0.005 \\
$\delta [\mathrm{rad}]$ & $-1.24^{+0.19}_{-0.05}$ & $-1.23^{+0.16}_{-0.05}$ & 0.008 & $-1.20^{+0.21}_{-0.08}$ & $-1.20^{+0.22}_{-0.08}$ & 0.004 \\
$d_{L} [\mathrm{Mpc}]$ & $507.5^{+128.6}_{-134.1}$ & $469.7^{+130.3}_{-124.8}$ & 0.043 & $504.8^{+115.0}_{-144.2}$ & $502.0^{+117.0}_{-142.1}$ & 0.003 \\
$\psi [\mathrm{rad}]$ & $1.45^{+1.50}_{-1.27}$ & $1.50^{+1.43}_{-1.29}$ & 0.003 & $1.41^{+1.52}_{-1.23}$ & $1.48^{+1.47}_{-1.29}$ & 0.009 \\
\hline 
	$N_s$ & $5476$ & $5162$ & - & $5093$ & $5027$ & -\\
	$T_s [h]$ & $63.7$ & $16.1$ & - & $18.5$ & $5.8$ & -\\
	$\ln\mathcal{Z}$ & $261.2$ & $261.6$ & - & $261.4$ & $261.3$ & - \\
\end{tabular}
\caption{
	Summary of parameter constraints for GW150914. The table compares the results from the \ac{FFD} and $\fs$ (Fs) methods for the two sampler settings (`s1' and `s2'). For each parameter, we report the median and the $90\%$ credible interval. The \ac{JSD} quantifies the consistency between the posteriors obtained from the `s1' and `s2' configurations for each method. The final three rows report summary statistics for each run: the number of posterior samples ($N_s$), the total sampling time in hours ($T_s$), and the log-evidence ($\ln\mathcal{Z}$).
}\label{tab:params}
\end{ruledtabular}
\begin{tablenotes}
    \small
    \item[a] 1. Parameter definitions follow the conventions used in the {\sc PESummary} package~\cite{Hoy:2020vys}.
    \item[b] 2. The geocenter time $t_c$ is reported as an offset from the GW150914 trigger time, $t_0=1126259462.4$ s.
\end{tablenotes}
\end{table*}

We note that the constraints on some parameters, such as the chirp mass $\mathcal{M}$, are slightly broader for the $\fs$ method compared to the \ac{FFD} method. This behavior is consistent with findings in other high-dimensional analyses, such as studies of black hole ringdown with multiple tones~\citep{Wang:2024jlz}. In such complex, often multimodal parameter spaces, standard nested sampling can sometimes struggle to explore the full posterior, leading to artificially narrow or biased constraints~\citep{sampling_errors2018,Wang:2024jlz}. The reduced dimensionality of the $\fs$ search space appears to mitigate these issues. Therefore, the slightly wider posteriors from the $\fs$ method represent a more conservative and honest quantification of the true statistical uncertainty.

Finally, the $\fs$ method provides a substantial improvement in computational efficiency. As detailed in the final rows of Table~\ref{tab:params}, for the `s1' (`s2') configuration, the \ac{FFD} analysis required $63.7$ ($16.1$) hours, while the $\fs$ analysis took only $18.5$ ($5.8$) hours. This corresponds to a speed-up factor of approximately $3.4$ ($2.8$), representing a reduction of about $70.6\%$ in total sampling time while producing a similar number of posterior samples.

\bibliographystyle{apsrev4-1}
\bibliography{fsIMR}

\end{document}